\documentclass[%
prereprint,
 amsmath,amssymb,
 aps,
 prfluids,
onecolumn,
longbibliography,unsortedaddress]{revtex4-2}

\usepackage{setspace}
\usepackage{booktabs}
\usepackage{mathrsfs}
\usepackage{graphicx}
\usepackage{dcolumn}
\usepackage{bm}
\usepackage{color}
\usepackage[x11names]{xcolor}

\usepackage{transparent} 
\usepackage{calc}		
\usepackage{soul} 		
	
\usepackage{float}		
\usepackage{subcaption}	
\graphicspath{{figures/}}	
\usepackage{comment}

\begin{document}
\let\vec\mathbf\ 

\title{The stability analysis of volatile liquid films in different evaporation regimes}
\author{Omair A. A. Mohamed}
\email{omair.mohamed@bilkent.edu.tr}
\affiliation{Mechanical Engineering Department, Bilkent University, Ankara, Turkey}

\author{Luca Biancofiore}
\email{\footnote{Corresponding author:}luca.biancofiore@univaq.it}
\affiliation{Department of Industrial Engineering, Information and Economics,
University of L’Aquila, Piazzale Ernesto Pontieri, Monteluco di Roio,
67100 L’Aquila, Italy}
\affiliation{Mechanical Engineering Department, Bilkent University, Ankara, Turkey}

\date{\today}

\begin{abstract}
	We investigate the role of the evaporation regime on the stability of a volatile liquid film flowing over an inclined heated surface using a two-fluid system that considers the dynamics of both the liquid phase and the diffusion of its vapor into the ambient environment.
	Consequently, the evaporation process is necessarily governed by the competition between (i) the thermodynamic disequilibrium tied to the liquid film's local thickness, and (ii) the diffusion effects dependent on the interface's curvature.
	We (i) modify the kinetic-diffusion evaporation model of Sultan \textit{et al.} [Sultan \textit{et al.}, J. Fluid Mech. \textbf{543}, 183, (2005)] to allow for the reduction in film thickness caused by evaporative mass loss, (ii) combine it with the liquid film formulation of Joo \textit{et al.} [Joo \textit{et al.}, J. Fluid Mech. \textbf{230}, 117, (1991)], and then (iii) utilize long-wave theory to derive a governing equation encapsulating the effects of inertia, hydrostatic pressure, surface tension, thermocapillarity, and evaporation.
	We utilize linear stability theory to derive the system's dispersion relationship, in which the Marangoni effect has two distinct components. 
	%
	The first results from surface tension gradients driven by the uneven heating of the liquid interface and is always destabilizing, while the second arises from surface tension gradients caused by imbalances in its latent cooling tied to vapor diffusion above it, and is either stabilizing or destabilizing depending on the evaporation regime. 
	These two components interact with evaporative mass loss and vapor recoil in a rich and dynamic manner.
	%
	Moreover, we identify an evaporation regime where the kinetic and diffusion phenomena are precisely balanced, resulting in a volatile film that is devoid of the vapor recoil and mass loss instabilities.
	%
	Additionally, we clarify the dependence of the mass loss instability on the wave number under the two-fluid forumlation, which we attribute to the presence of a variable vapor gradient above the liquid's surface.
	Furthermore, we investigate the effect of film thinning on its stability at the two opposing limits of the evaporation regime, where we find its impact in the diffusion-limited regime to be dependent on the intensity of evaporative phenomena.
	Finally, we conduct a spatiotemporal analysis which indicates that the strength of vapor diffusion effects is generally correlated with a shift towards absolute instability, while the thinning of the film is observed it to cause convective-to-absolute-to-convective transitions under certain conditions.
\end{abstract}
\maketitle

\newpage
\section{Introduction}
Liquid film evaporation is a process that is integral to numerous industrial applications, which include, among many others,  
the cooling of electronics \cite{1705142,KABOV2011825},
deposition-based production techniques \cite{Pütz2003,CHAURIS2015492,AhmadianYazdi2021},
thermal desalination systems \cite{AbrahamDesalination2013,QI2016207,JAMIL201876}, 
air humidifying devices \cite{HUANG2022125156}, 
absorption cooling cycles \cite{DOMINGUEZINZUNZA20161105, AMARIS2018826}, 
the cooling of solar panels \cite{KRAUTER2004131},
and solar-powered steam generation \cite{STRUCHALIN2020119987}.
These technologies are highly energy-intensive, and therefore, gaining a better understanding of the behavior of these liquid films carries the potential for significant environmental and economic gains, a goal that is at the core of the global effort to combat climate change.
In particular, investigating the spatiotemporal stability of these films by can help elucidate the complex interactions between the various physical phenomena governing their evolution. 
The value of such studies is heightened by the additional insight they can provide, such as the role that evaporation plays in the liquid film's transition into the turbulent flow regime, which has been linked to the film's thermal properties in addition to the Reynolds number \cite{chunseban1971}.
Investigations by means of stability analysis can help better comprehend such transitions, as some finite turbulent structures have been shown to emerge from specific infinitesimal perturbations \cite{lucas2017layer}.
The problem of an evaporating liquid film is a sophisticated and multifaceted challenge bringing together an array of physical phenomena which include inertia, thermocapillarity, phase change, thermodynamics, mass transfer, and diffusion. This complexity has provoked the development of various empirical and mathematical modeling techniques.

The initial efforts to study the stability of thin liquid films go back to the pioneering works of Kapitza \cite{kapitza1948wave,kapitza1949experimental}, who was the first to identify the existence of an inertial instability mode, now known as the Kapitza mode or \textit{H}-mode instability. 
The novel efforts of Kapitza on isothermal films were succeeded by studies by Benjamin \cite{benjamin1957wave} and Yih \cite{yih1963stability}, who determined the critical Reynolds number beyond which a vertical film is always unstable. 
A major advancement was later made by Benney \cite{benney1966long} who devised a lubrication-inspired approach to the problem which exploited the film's small vertical scale and its slow variation in space and time, a methodology now referred to as the long-wave expansion (LWE). 
The applicability of this approach was expanded substantially by Ruyer-Quil \& Manneville \cite{ruyer2000improved,ruyer2002further} by combining a boundary layer approximation with a projection of the velocity field onto an appropriate set of test functions, with the resulting equations being averaged using weighted residuals. This procedure, denoted as the weighted-residual integral boundary layer (WIBL) method, frees the flow's variables from being exclusively slaved to the film thickness, and avoids the unphysical blow up of Benney's \cite{benney1966long} technique at moderate Reynolds numbers \cite{pumir1983solitary}.
The correct mechanism behind the \textit{H}-mode instability was identified correctly as inertia by Smith \cite{smith1990mechanism}, with further clarifications about the role of inertia being made by Ruyer-Quil and Manneville \cite{RuyerQuil_Manneville_2005}, as well as Dietze \cite{dietze2016kapitza}.

Thermocapillary effects were introduced into the stability analysis of a liquid film by Sreenivasan and Lin \cite{sreenivasan1978surface} who considered a film flowing over an inclined substrate held at a fixed temperature, and detected a nonzero critical Marangoni number due to the stabilizing action of gravity. 
LWE was then applied to the problem by Kelly \textit{et al.} \cite{kelly1986instability} and yielded two critical Reynolds numbers corresponding to a thermocapillary instability  as well as a modified \textit{H}-mode instability.
Goussis and Kelly \cite{goussis1991surface} further classified this thermocapillary instability according to the responsible physical mechanism and the resulting form of motion into (i) the long-wave \textit{S}-mode instability whose horizontal length is much larger than the film's vertical scale, and (ii) the short wave \textit{P}-mode instability which produces patterns on the same scale as the film's thickness. 
Ruyer-Quil \textit{et al.} \cite{RyerThermo_part1_2005} examined long waves in a heated liquid film by applying the WIBL methodology to arrive at coupled equations for the evolution of the film thickness, the flow rate, and the interfacial temperature. Their system of equations was then analyzed by Scheid \textit{et al.} \cite{ScheidThermo_part2_2005} where good agreement was obtained with the Orr-Sommerfeld system, and the advective transport of heat was found to stabilize small-amplitude waves but destabilize large-amplitude waves in the presence of a recirculating flow region.

The impact of evaporation on the stability of liquid films and its relation to thermocapillarity has been studied extensively, both experimentally and theoretically.
Berg \textit{et al.} \cite{Berg_Boudart_Acrivos_1966} studied the formation of convection patterns in pools of evaporating liquids where the perceived patterns in many liquids were found to mostly depend on the depth of the liquid pool, as opposed to the properties of the liquid. 
Phase change can also cause the appearance of structures such as festoon instabilities, which were reported near the contact line of droplets by Redon \textit{et al.} \cite{festoonRedon1992}, as well as height fluctuations as observed by Kavehpour \textit{et al.} \cite{KAVEHPOUR2002409} along the droplet's entire surface. 
Kimball \textit{et al.} \cite{Kimball_Evap_2012} investigated the evolution of convective structures in thin liquid films where they observed a sequence of transitions as the film thins from large variable cells, to concentric rings and spirals, to the eventual cessation of Marangoni convection. Their analysis suggests that attaining equilibrium conditions at the liquid's surface unifies its temperature and ultimately causes the elimination of the thermocapillary instability.
The phenomenon of the tears of wine was studied extensively by means by Fournier and Cazabat \cite{JBFournier1992}, Vuilleumier \textit{et al.} \cite{Vuilleumier1995}, and Hosoi and Bush \cite{HOSOI_BUSH_2001}.
Moreover, the influence of non-uniform heating was studied by Mildinova \textit{et al.} \cite{MILADINOVA_SLAVTCHEV_LEBON_LEGROS_2002}, Kalliadasis \textit{et al.} \cite{KALLIADASIS_KIYASHKO_DEMEKHIN_2003}, and Yeo \textit{et al.} \cite{PhysRevE.67.056315}.

Formulations of the evaporation process which do not account for the dynamics of the vapor phase beyond the interfacial boundary conditions are known as \textit{one-sided} models. They represent scenarios where the liquid is evaporating into a vacuum or subjected to continuous gas flow \cite{mohamed_luca_21}.
These models treat the evaporation process as a departure from thermodynamic equilibrium at the liquid interface, and intrinsically predict the effects of evaporation to be destabilizing, as argued by Prosperetti and Plesset \cite{Prosperetti1984}. The imbalance of pressure exerted by the liquid molecules departing the interface produces an effect known as vapor recoil, which was first reported experimentally by Hickman \cite{Hickman1952} in flat liquid films under high vacuum, who found it to both strongly destabilize the interface and dramatically increase the evaporation rate. This one-sided problem was the subject of repeated theoretical efforts such as that undertaken by Palmer \cite{palmer1976hydrodynamic}, who investigated the coupling between the Marangoni and vapor recoil instabilities for long-wave disturbances. 
Similar studies followed by Burelbach \textit{et al.} \cite{burelbach1988nonlinear} who considered a flat film and included the effects of evaporative mass loss and disjoining pressure, Danov \textit{et al.} \cite{Danov1998} who accounted for the presence of nonvolatile surfactants, and Joo \textit{et al.} \cite{joo1991} who analyzed an inclined film and incorporated higher order effects. 
More recently, Mohamed and Biancoﬁore \cite{mohamed2020linear} circumvented the limitations imposed by long-wave theory on the model of Burelbach \textit{et al.} \cite{burelbach1988nonlinear} by deriving the corresponding Orr-Sommerfeld system instead. We follow them herein and adopt the terminology \textit{E}-mode to refer to the evaporative instability.
Subsequent investigations into this problem were conducted by by Oron \textit{et al.} \cite{oron1997long}, Margerit \textit{et al.} \cite{Margerit2003}, and Merkt and Bestehorn \cite{MERKT2003196}. 
Additional efforts were made to further develop one-sided models such as those exerted by Ajaev and Homsy \cite{ajaev2002steady,ajaev2005spreading,ajaev2005evolution} who applied their model to investigate the effect of evaporation on wetting phenomena.
Ji and Witelski \cite{ji2018instability} examined the competition between weak phase change and dewetting in thin liquid films, and observed complex evolution dynamics.
%
Notable pursuits were also made to improve the modeling of the thermodynamic balance at the liquid interface such as that conducted by Shklyaev and Fried \cite{shklyaev2007stability} who wrote a novel set of boundary conditions to replace the conventional Hertz-Knudsen-Langmuir equation with a formulation based on the balance of conﬁgurational momentum, which showed realistic applicability to molten metals.

On the other hand, in the absence of sufficient advection of the liquid's vapor from the vicinity of its surface, vapor diffusion becomes an important effect producing intricate phenomena and therefore must be taken into consideration. Deegan \textit{et al.} \cite{deegan1997capillary} explained the origin of coffee stains in the context of the diffusion-limited evaporation of droplets into ambient air, while Cachile \textit{et al.} \cite{Cachile2002} explained the receding of unpinned droplets under similar assumptions. Poulard \textit{et al.} \cite{Poulard2003} studied a similar problem where contact line instabilities and the nonmonotonic behavior of the receding contact line were explained. 
These instabilities inspired an important development in the modeling of evaporating liquid films made by Sultan \textit{et al.} \cite{sultan2005evaporation}, who formulated a general evaporation model which incorporates both the thermodynamically determined molecule departure rate as well their diffusion into the ambient gas phase, under the assumption of the static diffusion of the vapor. 
This assumption was relaxed by Kanatani \cite{Kanatani_2013}, who investigated a liquid film evaporation into an inert gas where the convection of the vapor was taken into account. Additional progress was made by Zhao and Nadal \cite{Zhao_Nadal_2023}, who addressed the Stefan ﬂow, the temperature jump across the Knudsen layer, and the change in the interface's chemical potential due to the presence of the inert gas.


In this work we study the influence of the evaporation regime on the Marangoni instability in the two-fluid problem of a volatile liquid film flowing down an inclined plane while its vapor diffuses into the surrounding gas phase. The presence of vapor diffusion gives rises to complex dynamics resulting from the interplay between the Marangoni instability and evaporation phenomena such vapor recoil, mass loss, and film thinning. These interactions have yet to be thoroughly investigated, and hence, the goal of this manuscript is to fill this knowledge gap by means of the temporal and spatiotemporal analyses of the two-fluid system. 
Our analysis elucidates the dependence of the mass loss instability on the wave number in the presence of vapor diffusion, which we attribute to the nonuniform vapor gradient above the liquid's surface.
In addition, we explore the influence of the evaporation regime on the effects resulting from film thinning as well as wave celerity within the liquid film. 
%

This manuscript is structured as follows.
In Sec. \ref{problem_formulation}, we introduce the equations governing both the liquid and vapor domains, as well as the scaling employed, and the resulting set of dimensionless parameters. The section is then concluded with our application of the LWE to produce the film's governing equation, which is perturbed to produce the dispersion relationship representing the system's dynamic behavior. 
In Sec. \ref{sec_temporal_stability}, the dispersion relationship is utilized to perform a temporal stability analysis where we probe the influence of the evaporation regime on the film's temporal stability, the mass loss instability, the effects resulting from film thinning, and wave celerity.
The temporal analysis is expanded into the spatiotemporal framework in Sec. \ref{spatiomteporal_stability}, where we examine the impact of the evaporation regime as well as film thinning on the spatiotemporal stability. Finally, our summary and conclusions are presented in Sec. \ref{conclusion}.

\color{black}

\section{Theoretical formulation}\label{problem_formulation}
\begin{figure}[tp]
	\captionsetup{justification=raggedright}
	\begin{center}
		\scalebox{0.6}{\includegraphics{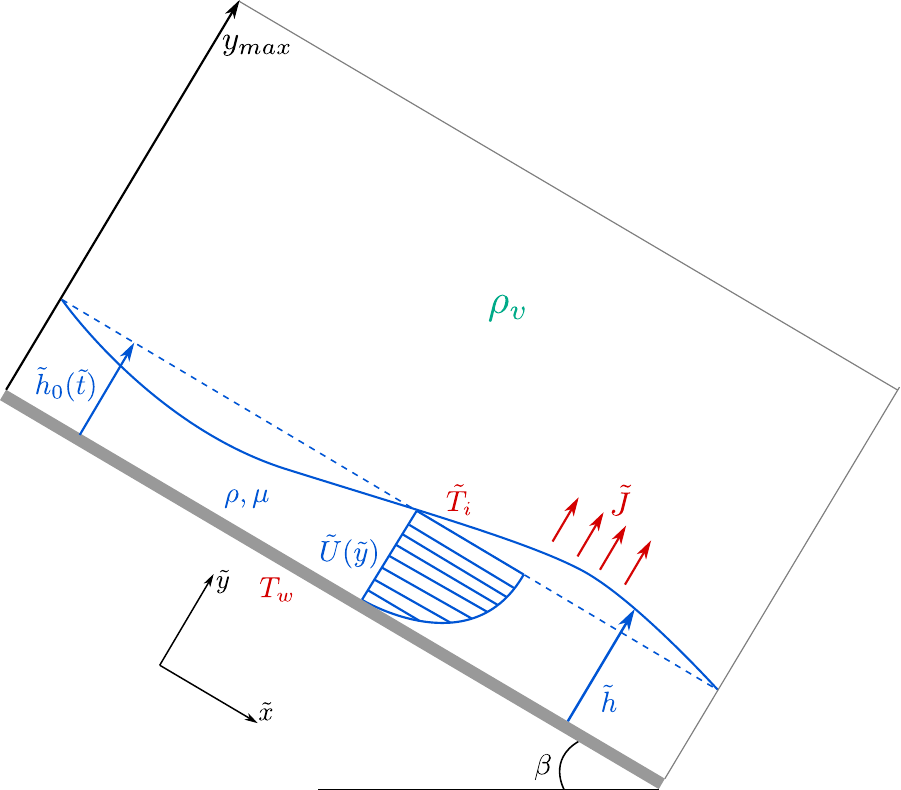}}
	\end{center}
	\caption{Schematic illustration of an evaporating liquid film flowing under the influence of gravity over a heated plate held at a constant temperature. The unperturbed base flow height $\tilde{\bar{h}}$ is marked by the dashed line, $\beta$ denotes the inclination angle from the horizontal, and $T_w$ is the constant wall temperature.}
	\label{liquid_film_schematic}
\end{figure}

We study a two-phase problem combining the studies of Joo \textit{et al.} \cite{joo1991} and Sultan \textit{et al.} \cite{sultan2005evaporation} where we investigate a flowing liquid film evaporating into an adjacent gas phase. The fluid film is incompressible, thin, Newtonian, volatile, with constant physical properties, and driven by gravity down a heated plate held at a constant temperature, which is inclined from the horizontal with an angle $ \beta $, as depicted in Fig. \ref{liquid_film_schematic}.

Our study is limited to two dimensions where we assign Cartesian $ (\tilde{x},\tilde{y}) $ coordinates such that the $ \tilde{x} $ coordinate points in the streamwise direction and the $ \tilde{y }$ coordinate away from the liquid film into the gas phase. The function $\tilde{h}(\tilde{x},\tilde{t})$ defines the liquid interface forming the boundary between the two phases, while $\tilde{y}_{max}$ marks the upper limit of the gas domain. The gas phase consists of a static mixture of an inert gas and the liquid's vapor and does not contribute any viscous stresses at the liquid interface. Note that the tilde notation denotes dimensional variables.
Our focus in this investigation is on the thermal instability mode appearing at relatively low Reynolds numbers and which results from the coupling of the thermocapillary \textit{S}-mode \cite{goussis1991surface} and the evaporative \textit{E}-mode \cite{mohamed2020linear} instabilities.
\subsection{Liquid phase governing equations}
The equations governing the liquid film are the continuity equation, the Navier-Stokes momentum equations, and the energy equation, which in dimensional vector form read
\begin{subequations}\label{governing_dimensional_vectorform}
\begin{align}
	&\nabla\cdot\vec{\tilde{u}} = 0,
	\\ &\partial_{\tilde{t}} \vec{\tilde{u}} + \vec{\tilde{u}}\cdot\nabla\vec{\tilde{u}} = -\frac{1}{\rho}\nabla {\tilde{p}} + \nu\nabla^2\vec{\tilde{u}}+\vec{\tilde{F}},
	\\ &\partial_{\tilde{t}} \tilde{T} + \vec{\tilde{u}}\cdot\nabla \tilde{T} = \alpha\nabla^2\tilde{T}. 
\end{align}
$\vec{\tilde{u}} = (\tilde{u}, \tilde{v})$  is the liquid velocity vector,
$ \tilde{p} $ is the liquid's pressure,
$ \tilde{T} $ is the liquid temperature, 
$ \vec{\tilde{F}} = (g\sin\beta,\ g\cos\beta)$ is the gravitational body force vector,
$ \rho $ is the liquid density,
$ \nu $ is the liquid's kinematic viscosity, 
and $ \alpha  = \lambda / \rho c_p$ is the liquid's thermal diffusivity involving liquid's thermal conductivity $\lambda$, and its constant pressure heat capacity $c_p$.

At the solid wall $( \tilde{y} = 0 )$, we impose a no slip condition on the velocity and a constant value for the temperature:
\begin{align}
	&\vec{\tilde{u}} = 0,
	\\ &\tilde{T} = T_{wall}.
\end{align}
At the interface between the liquid and the gas $\tilde{y} = \tilde{h}(\tilde{x},\tilde{t})$ the boundary conditions read
\begin{align}
	&\tilde{J} = \tilde{\rho}\left(\vec{\tilde{u}}- \vec{\tilde{u}}_i\right)\cdot\vec{n} = \tilde{\rho}_v\left(\vec{\tilde{u}}_v- \vec{\tilde{u}}_i\right)\cdot \vec{n}, 
	\\ &\tilde{J}\left[\mathcal{L} + \frac{1}{2}\left(\frac{\tilde{J}}{\tilde{\rho}_v}\right)^2\right]= -\lambda \nabla \tilde{T}\cdot\vec{n},\label{dimensional_energy_balance}
	\\ &\frac{\tilde{J}^2}{\tilde{\rho}_v}\vec{n}+(\vec{\tilde{T}}-\vec{\tilde{T}}_g)\cdot\vec{n} =2\tilde{\gamma} \mathcal{K}\vec{n}+\nabla_s\tilde{\gamma}.\label{general_stress_balance}
\end{align}
The subscript $ i $ denotes quantities which are at the liquid-gas interface, the subscript $ v $ denotes quantities associated with the liquid's vapor, and the subscript $ g $ denotes those representing the ambient gas phase. 
$ \mathcal{L} $ is the latent heat of vaporization and $ \tilde{J} $ is the vapor mass flux.
Moreover, $\mathcal{K} = -\frac{1}{2}\nabla_s\cdot\vec{n}$ is the mean interface curvature, 
$\nabla_s$ is the surface gradient operator, and $ \vec{n} = \frac{1}{N}(-\partial_{\tilde{x}} \tilde{h},1) $ is the local interface normal unit vector.
The stress tensor for the liquid phase is $ \vec{\tilde{T}} = -\tilde{p}\vec{I} + \vec{\tilde{P}}$ where $ \vec{\tilde{P}} $ is the deviatoric stress tensor, while the gas phase's stress tensor is simplified to $\tilde{\vec{T}}_g = -\tilde{p}_g \vec{I}$. 
Furthermore, the liquid's surface tension is defined by a linear equation of state
\begin{align*}
	\tilde{\gamma} = \gamma_0 - \left|\dfrac{d\tilde{\gamma}}{d\tilde{T}}\right|(\tilde{T} - T_{wall}),
\end{align*} 
where $ \gamma_0 $ is the the mean surface tension at $ T_{wall}$ and $ d\tilde{\gamma}/d\tilde{T}$ is negative for most liquids.
\end{subequations}
\subsection{Vapor phase governing equations}
The vapor phase is modeled by modifying Sultan \textit{et al.}'s \cite{sultan2005evaporation} formulation for the liquid vapor to accommodate the thinning of the film due to evaporative mass loss.
Under the assumption of quasi-static diffusion, the P\'eclet number is very small and the convection-diffusion equation governing the liquid vapor distribution reduces to the Laplace equation:
\begin{subequations}\label{vapor_system_dimensional}
	\begin{align}
	\dfrac{\partial^2 \tilde{\rho}_v}{\partial \tilde{x}^2} + \dfrac{\partial^2 \tilde{\rho}_v}{\partial \tilde{y}^2} = 0.
\end{align}
The interface boundary condition is derived by imposing the continuity of the normal vapor flux at liquid interface, thereby equating the vapor flux required by the Hertz-Knudsen relationship \cite{Prosperetti1984} to that required by Fick's diffusion law, to obtain at $\tilde{y} = \tilde{h}$:
\begin{align}
	-D\Big(1 + v_{th} &\dfrac{d\tilde{\rho}_v^{eq}}{d\tilde{T}}\dfrac{\mathcal{L}}{\lambda}\tilde{h}\Big)(\vec{n}\cdot\nabla)\tilde{\rho}_{v} 
	= v_{th}\Big(\tilde{\rho}_v^{eq}(T_{wall}) - \tilde{\rho}_{v}\Big),
\end{align}
with the vapor mass flux being
\begin{align}
	\tilde{J} = -D(\vec{n}\cdot \nabla)\tilde{\rho}_{v|int}. \label{flux_expression}
\end{align}
$D$ is the diffusion coefficient of vapor into the gas phase, $\tilde{\rho}_v^{eq}$ is the equilibrium vapor density, $\tilde{\rho}_{v|int}$ is the vapor density at the liquid interface $\tilde{h}(\tilde{x},\tilde{t})$, and $v_{th} = \alpha_M\sqrt{k_B \tilde{T}_{int}/(2\pi M_w)}$ is the typical thermal kinetic velocity with the accommodation coefficient $\alpha_M$, the Boltzmann constant $k_B$, and the molecular weight $M_w$. The detailed derivation of this boundary condition can be found in Ref. \cite{sultan2005evaporation}. 
Moreover, to prevent the saturation of the gas domain with vapor, a vapor outflux is enforced at the upper boundary by imposing the density gradient. 
We write a boundary condition that is tied to the liquid film's height, which allows the vapor density gradient to react as the film becomes thinner due to the evaporative mass loss. Moreover, the boundary condition is formulated to recover a single-fluid model when the appropriate limit of the two-phase system is taken. In particular, our expression allows the recovery of the exact expressions in Joo \textit{et al.} \cite{joo1991}, who considered only the liquid phase in their work. The top boundary condition at $\tilde{y} = y_{max}$ reads:
\begin{align}
	\dfrac{\partial \tilde{\rho}_v}{\partial \tilde{y}} = 
	-\dfrac{v_{th}\left(\tilde{\rho}^{eq}_{v|Twall} - \tilde{\rho}^{eq}_{v|Tint}\right)}
	{-D\left(\vec{n}\cdot\nabla\right)\tilde{\rho}_{v,int} + v_{th}\left(\tilde{\rho}^{eq}_{v|Twall} - \tilde{\rho}^{eq}_{v|Tint}\right)}
	\dfrac{\lambda\Delta T}{\mathcal{L}D}\dfrac{1}{\tilde{h}},
\end{align}
where the fraction represents the degree of thermodynamic disequilibrium at the interface allowing the recovery of the one-sided model, and $\lambda\Delta T/(\mathcal{L}D)$ is the characteristic vapor density.
The formulation of this boundary condition represents an extension of the model of Sultan \textit{et al.} \cite{sultan2005evaporation} to non-equilibrium evaporation and allows the inclusion of the resultant thinning of the film due to the evaporative mass loss, as their model is restricted to recovering the one-sided case of a film at full height undergoing equilibrium evaporation where net mass loss is zero. 
Finally, the system of equations governing the vapor phase is closed by periodic boundary conditions at the horizontal extremities of the domain to facilitate its numerical solution:
\begin{align}
	\tilde{\rho}_v(x_0,\tilde{y}) = \tilde{\rho}_v(x_{max},\tilde{y}),\quad \left.\dfrac{\partial \tilde{\rho}_v}{\partial\tilde{x}}\right|_{x_0} = \left.\dfrac{\partial \tilde{\rho}_v}{\partial\tilde{x}}\right|_{x_{max}},
\end{align}
\color{black}
where $x_0$ and $x_{max}$ are the horizontal domain limits.
\end{subequations}


\subsection{Scaling and dimensionless parameters}
The following scales are introduced into the systems of equations \eqref{governing_dimensional_vectorform} and \eqref{vapor_system_dimensional}: 
\begin{gather}\label{scaling}
	\begin{align*}
		&(\tilde{x},\tilde{y},\tilde{z}) \rightarrow \tilde{\bar{h}}_0(x^*,y^*,z^*), 	&  &\tilde{h} \rightarrow  \tilde{\bar{h}}_0 h^*, & 	&\tilde{t} \rightarrow \frac{\tilde{\bar{h}}_0^2}{\nu}t^*, & & \tilde{T} \rightarrow T_w + T^*\Delta T, & \\
		& 	(\tilde{u},\tilde{v},\tilde{w}) \rightarrow \frac{\nu}{\tilde{\bar{h}}_0}(u^*,v^*,w^*),	&  &\tilde{p} \rightarrow p_g + \frac{\rho\nu^2}{\tilde{\bar{h}}_0^2}p^*,         & 	&\tilde{J} \rightarrow \frac{\lambda \Delta T}{\tilde{\bar{h}}_0\mathcal{L}}J^*, & & \tilde{\rho}_v \rightarrow \rho^{eq}_v(T_{wall}) +\dfrac{\lambda\Delta T}{\mathcal{L}D}\rho_v^*,
	\end{align*}
\end{gather}
where $ \Delta T = T_w-T_g $, and $ \tilde{\bar{h}}_0 $ is the flat film thickness at the initial time. The star notation denotes dimensionless variables, and is dropped onwards for simplicity. 
Introducing these scales into Eqs. \eqref{governing_dimensional_vectorform} and \eqref{vapor_system_dimensional} yields the liquid film equations as
\begin{subequations}\label{governing_dimensionless_2D}
	\begin{align}
	&\partial_{x} u + \partial_{y} v = 0 , 
	\\&\partial_{t} u  + u \partial_{x} u + v \partial_{y} u =  -\partial_{x} p + \partial_{xx} u  + \partial_{yy} u  + Re\sin\beta, 
	\\&\partial_{t} v  + u \partial_{x} v + v \partial_{y} v =  -\partial_{y} p + \partial_{xx} v  + \partial_{yy} v - Re\cos\beta,  
	\\&Pr\left[\partial_{t} T  + u \partial_{x} T + v \partial_{y} T\right] = \partial_{xx} T + \partial_{yy}T.\label{energy_eqn_ND}
	\end{align}
	The scaled wall	boundary conditions at $ (y = 0) $ are
	\begin{align}
	& u = v = 0,\qquad T = 0, \label{wall_BC_ND}
	\end{align} 
	while the scaled interface	boundary conditions at $ (y = h) $ become
	\begin{align}
	&EJ = \frac{1}{N}\left[-u\partial_{x}h + v -\partial_{t}h\right],\label{kinematic_BC}
	\\&J + \frac{E^2}{D^2L}{J}^3 = \frac{1}{N}\left[\partial_{x}h\partial_{x}T - \partial_{y}T\right],\label{energy_balance}
	\\\begin{split}
	&-\frac{3}{2}\frac{E^2}{D}{J}^2 - \frac{2}{N^2}\left[\partial_{x}u\left[(\partial_{x}h)^2 - 1\right] 	-\partial_{x}h(\partial_{y}u + \partial_{x}v) \right] + p = 	-3S(1-CT)\left[\frac{\partial_{xx}h}{N^3}\right],
	\end{split}
	\\&\left[1-(\partial_{x}h)^2\right][\partial_{y}u+\partial_{x}v] - 4\partial_{x}h\partial_{x}u -\tau = -2\frac{M}{P}\left[\partial_{x}T + \partial_{x}h\partial_{y}T\right]N.
	\end{align} 
\end{subequations}
Similarly, the dimensionless vapor system is
\begin{subequations}\label{vapor_system_ND}
	\begin{align}
		&\dfrac{\partial^2 \rho_v}{\partial x^2} + \dfrac{\partial^2 \rho_v}{\partial y^2} = 0, \\[0.15cm]
		&(1 + \chi h)(\vec{n}\cdot \nabla)\rho_{v} = Pe_k \rho_{v}\qquad \text{at } y = h, \label{BC_vapor_ND} \\[0.15cm]
		&\dfrac{\partial \rho_v}{\partial y} = -\dfrac{\chi}{1 + \chi h}\qquad  \text{at } y = y_{max}, \\[0.15cm]
		&\rho_v(x_0,y) = \rho_v(x_{max},y), \quad \left.\dfrac{\partial \rho_v}{\partial x}\right|_{x_0} = \left.\dfrac{\partial \rho_v}{\partial x}\right|_{x_{max}}
	\end{align}
with the dimensionless expression for the vapor mass flux obtained from Eq. \eqref{flux_expression} as
\begin{align}
	J = -(\vec{n}\cdot \nabla)\rho_{v|int}. \label{flux_expression_dimensionless}
\end{align}
\end{subequations}

Equation \eqref{BC_vapor_ND} is identical to the expression in Sultan \textit{et al.}, \cite{sultan2005evaporation} and contains the two parameters determining the evaporation regime: the thermal expansion number $\chi$ representing the fluctuations in the equilibrium  vapor density with temperature, and the \textit{kinetic} P\'eclet number $Pe_k$, which is the ratio of kinetic and diffusion time scales. The ratio $Pe_k/\chi$ which we denote as $\Gamma $ defines the evaporation regime. 
When $\Gamma < 1$ the evaporation rate is dictated to a greater degree by the thermodynamic molecular exchange at the liquid interface, and is, thus, strongly dependent on the interface's proximity to the heated wall as opposed to its curvature
. Consequently, the evaporation rate will be higher (lower) at the film's troughs (crests).
The particular case of $\Gamma \rightarrow 0$ represents the exclusion of liquid vapor as a variable in the system, and hence, the complete absence of vapor diffusion phenomena. In this situation, the problem is reduced to a single-fluid (one-sided) system where the evaporation process is slaved to the local film height, since it dictates the energy transfer to the the interface and the resulting local vapor mass flux. This evaporation regime is therefore aptly named the transfer-rate-limited regime, and is the evaporation regime considered by several of the notable works investigating the stability of volatile films \cite{burelbach1988nonlinear,joo1991,Margerit2003,ajaev2002steady,ajaev2005evolution,ajaev2005spreading,ji2018instability,mohamed_luca_21}.

On the other hand, when $\Gamma >1$, the gas layer directly above the liquid is sufficiently saturated with vapor such that the evaporation rate is  largely dependent on the diffusion of vapor into the ambient phase, such that kinetic phenomena take a secondary role. Consequently, the evaporation rate is determined by the interface's curvature since it governs the vapor gradient directly above the liquid film. Under these conditions, the evaporation rate will be higher (lower) at the film's crests (troughs) due to their concave (convex) shape, which allows for higher overall vapor diffusion, in what is known as the sharp-edge effect. For more details please refer to Refs. \cite{deegan1997capillary,deegan2000contact,sultan2005evaporation}.
Moreover, when $\Gamma \rightarrow \infty$, the layer of gas directly above the liquid surface is completely saturated by vapor and the evaporation process is entirely dependent on the diffusion rate. Thus, the evaporation conditions at this threshold are known as the diffusion-limited evaporation regime \cite{sultan2004diffusion,sultan2005evaporation}.
\color{black}
In this work we investigate both transfer-rate-limited and diffusion-limited evaporation, as well as intermediate cases where kinetic and diffusion processes compete at the liquid interface.

\color{black}
\begin{table*}
	\caption{The set of dimensionless parameters resulting from the scaling of Eqs. \eqref{governing_dimensional_vectorform} and \eqref{vapor_system_dimensional}.}
	\begin{ruledtabular}\label{dimensionless_parameters}
			\begin{tabular}{l c l} 
				Parameter			 & Expression					 & Physical significance \\ \hline
				Reynolds number $ (Re) $ 			 				 & $ g\tilde{\bar{h}}_0^3/\nu^2 $ 				    & Ratio of inertial to viscous forces			\\
				Prandtl number $ (Pr) $				 				 & $ \nu/\alpha $								& Ratio of momentum to thermal diffusivity. 		\\
				Evaporation number $ (E) $			 				 & $ \lambda\Delta T/\rho \mathcal{L}\nu $			& Ratio of viscous to evaporative timescales 	\\
				Density ratio $ (D) $				 				 & $ 3\tilde{\rho}_v/2\tilde{\rho} $							& Ratio of vapor density to liquid density.		\\
				Non-dimensional latent heat $ (L) $	 				 & $ 8\tilde{\bar{h}}_0^2\mathcal{L}/9\nu^2 $		& Measure of the liquid's latent heat.			\\
				Non-dimensional surface tension $ (S) $				 & $\gamma_0\tilde{\bar{h}}_0/3\tilde{\rho}\nu^2 $					&  Measure of the liquid's surface tension.   	\\
				Capillary number $ (C) $				 			 & $ \gamma\Delta T/\gamma_0 $				& Ratio of viscous to surface tension forces.	\\
				Marangoni number $ (M) $							 & $|d\tilde{\gamma}/d\tilde{T}|\Delta T\tilde{\bar{h}}_0/2\mu\alpha $		& Ratio of Marangoni to viscous forces.	        \\
				Thermal expansion parameter $ (\chi) $				 & $v_{th}\tilde{\bar{h}}_0(\mathcal{L}/\lambda) (d\rho_v^{eq}/dT) $		& Fluctuations in vapor equilibrium density.	        \\
				Kinetic P\'eclet number $ (Pe_k) $					 & $v_{th}\tilde{\bar{h}}_0/D $		& Ratio of kinetic and diffusion time scales.	        \\
				Evaporation regime parameter $(\Gamma) $             & $\lambda/[D(d\rho_v^{eq}/dT)\mathcal{L}]$ 		& Relative importance of kinetic and diffusion effects.	        \\
			\end{tabular}
	\end{ruledtabular} 
\end{table*}

\subsection{Long wave theory and the interface evolution equation}\label{LWEandGovEqn}
The long wave expansion (LWE) is employed by defining a small parameter $ \epsilon  = \dfrac{h}{\ell}$ where $ \ell $ is the flow's horizontal characteristic length and $ h \ll \ell   $. 
The liquid system's independent variables are then rescaled using $\epsilon$ as
\begin{gather}\label{rescaling_indep_variables}
	\begin{align}
	&x \rightarrow \epsilon x, & & y \rightarrow y, & & t \rightarrow \epsilon t. &
	\end{align}
\end{gather}
Now the system's dependent variables are expanded in terms of $\epsilon$ as
\begin{subequations}\label{dependent_scales}
	\begin{align}
		&u = u_0 + \epsilon u_1 + \mathcal{O}(\epsilon^2),\\
		&v = \epsilon(v_0 + \epsilon v_1 + \mathcal{O}(\epsilon^2)),\\		
		&p = p_0 + \epsilon p_1 + \mathcal{O}(\epsilon^2),\\
		&T = T_0 + \epsilon T_1 + \mathcal{O}(\epsilon^2).
	\end{align}
\end{subequations}
Additionally, the following dimensionless parameters are rescaled:
\begin{gather}\label{parameter_rescaling}
\begin{align}
\hskip1.5cm & E \rightarrow \epsilon\bar{E}, & & D \rightarrow \epsilon^2\bar{D}, & & S \rightarrow \dfrac{\bar{S}}{\epsilon^2}. &
\end{align}
\end{gather}
Note that the transformations in Eqs. \eqref{parameter_rescaling} can be considered realistic when compared to the data tabulated in Ref.  \cite{burelbach1988nonlinear} and that since $\mathcal{L}$ is typically quite large, the kinetic energy term in the energy balance at the liquid interface [Eq. \eqref{energy_balance}] can be neglected \cite{burelbach1988nonlinear,joo1991,oron1997long}. 
Moreover, we consider the vapor mass flux $(J)$ and the interface height $(h)$ functions to be leading order quantities.
Eqs. \eqref{rescaling_indep_variables}, \eqref{dependent_scales}, and \eqref{parameter_rescaling} are substituted into the system of equations \eqref{governing_dimensionless_2D} and the ordering procedure allows obtaining the leading temperature solution as 
\begin{align}
	T = -J(x,t)y.\label{leading_temperature_ND}
\end{align}
Equation. \eqref{leading_temperature_ND} is sufficient to substitute the vapor mass flux derivatives in place of the temperature derivatives in the $\mathcal{O}(\epsilon)$ system since our application of the LWE drops terms of order $\mathcal{O}(\epsilon^2)$.
The vapor mass flux is retained as an unknown function, and we obtain solutions for the dependent variables $ (u,v,p) $ which are then substituted into Eq. \eqref{kinematic_BC} to obtain the film evolution equation as
\begin{align}\label{Benney_equation} 
	\partial_t h + \bar{E}J + 
	\partial_x \Bigg\{&\dfrac{1}{3}Re\sin\beta h^3 +\epsilon\ \left[  \dfrac{2}{15}\left(Re\sin\beta\right)^2h^6 \partial_x{h} + \dfrac{5}{24}\bar{E}(Re\sin\beta) h^4J  \right. \nonumber
	\\& \left. - \dfrac{1}{3}(Re\cos\beta) h^3 \partial_x{h} +\mathcal{M}\partial_x\left(hJ\right)h^2 - V_r h^3J \partial_xJ + \bar{S}h^3 \partial_{xxx}h   \right] \Bigg\} =0. 
\end{align}
\color{black}
Equation \eqref{Benney_equation} describes the evolution of the liquid interface under the influence of gravity, hydrostatic pressure, thermocapillarity, and evaporation. Note that the exact governing equation obtained by Joo \textit{et al.} \cite{joo1991} can be recovered from this expression by substituting $J = 1/(h + K)$. 
Our expression represents an extension of the model of Sultan \textit{et al.} \cite{sultan2005evaporation} to include thermocapillary effects at $\mathcal{O}(1)$, as well as convection and evaporation phenomena at $\mathcal{O}(\epsilon)$. 
\color{black}
Moreover, we have defined $ V_r = \bar{E}^2/D$ as the parameter representing the strength of vapor recoil forces, as well as $ \mathcal{M} = M/Pr $, which represents the intensity of the Marangoni effect. 
Note that our employment of the LWE is justified here since its accuracy is comparable to more advanced models for Reynolds numbers up to $\mathcal{O}(10)$ \cite{vellingiri2015absolute}, which is consistent with our flow regime of interest, characterized by thermal instabilities appearing at low Reynolds numbers.
\color{black}

\subsection{Dispersion relationship derivation}\label{sec_dispersion_derivation}
The base states for film height, vapor mass flux, and vapor density can be found by setting $ \partial_x = 0 $ in Eqs. \eqref{vapor_system_ND}  and \eqref{Benney_equation} to arrive at the following base state solutions, respectively:
\begin{align}\label{base_states}
	 \bar{h}(t) = \sqrt{(\bar{h}_0 + \dfrac{1}{\chi})^2 - 2\bar{E}t} - \dfrac{1}{\chi},
	\qquad \bar{J} = \dfrac{1}{\bar{h}+ \dfrac{1}{\chi}},
	\qquad \bar{\rho}_v(t,y) = \bar{J}(\bar{h} - y) - \dfrac{1}{\Gamma},
\end{align}
where $\bar{h}_0$ is the initial dimensionless base flow height at $t = 0$. 
In contrast to Sultan \textit{et al.}'s \cite{sultan2005evaporation} formulation  in which $\bar{J} = 1$, our formulation results in the base density gradient and the associated base vapor flux depending on $\bar{h}(t)$, and therefore can account for the higher evaporation rate expected as the volatile film's thickness reduces.
Note that the expressions for $\bar{h}(t)$ and $\bar{J}$ in Eq. \eqref{base_states} are equivalent to those in Burelbach \textit{et al.} \cite{burelbach1988nonlinear} and Joo \textit{et al.} \cite{joo1991}, with $1/\chi$ taking the place of their interfacial non-equilibrium parameter $K$.

We now employ the assumption that the perturbation's growth is much faster than the variation in the base flow's height due to evaporation, i.e., that the base flow's height can be considered \textit{frozen} and hence treated as a steady state during perturbation growth \color{black} \cite{davis1976stability, burelbach1988nonlinear}. We introduce the following infinitesimal perturbations into the coupled system's variables:
	\begin{align}\label{perturbations1}
		h = \bar{h} + \delta h,
		\qquad \rho_v = \bar{\rho}_v + \delta\rho_v,
		\qquad J = \bar{J} + \delta J,
	\end{align}
where $\delta h = He^{i(kx - \omega t)}  + c.c.$, $H\ll1$, and c.c. stands for the complex conjugate. Due to the applied scaling [Eqs. \eqref{parameter_rescaling}], here $ k = k^*/\epsilon $ and $ \omega = \omega^*/\epsilon $, where $ k^* $ and $ \omega^* $ are the wavenumber and angular frequency in the nonscaled dimensionless spatial and temporal coordinates $ (x^*, t^*) $\color{black}. 
We then utilize Eq. \eqref{BC_vapor_ND} to express $\delta\rho_v$ in terms of $\delta h $
\begin{align}\label{perturbations2}
	\delta \rho_v = \bar{J}\dfrac{\Gamma - 1}{\Gamma + |k|\left(\bar{h} + \dfrac{1}{\chi} \right)}e^{-|k|(y - \bar{h})}\delta h,
\end{align}
and subsequently use Eq. \eqref{flux_expression_dimensionless} to finally express $\delta J$ in terms of $\delta h$
\begin{align}\label{perturbations3}
	 \delta J = \bar{J}|k|\dfrac{\Gamma - 1}{\Gamma + |k|\left(\bar{h} + \dfrac{1}{\chi} \right)}\delta h.
\end{align}
Substituting from Eqs. \eqref{perturbations2} and \eqref{perturbations3} into Eq. \eqref{Benney_equation} and linearizing yields an expression for the dispersion relationship governing the evolution dynamics 
\begin{align}\label{dispersion_relationship_general}
	\omega = &\left[Re\sin\beta\bar{h}^2  + \epsilon\dfrac{5}{6}Re\sin\beta\bar{E}\bar{J}\bar{h}^3\right]k \nonumber
	\\& \hskip0.75cm + i\epsilon\left[\dfrac{2}{15}\left(Re\sin\beta\right)^2 \bar{h}^6 - \dfrac{1}{3}Re\cos\beta\bar{h}^3 + \mathcal{M}\bar{J} \bar{h}^2\right]k^2 - i\epsilon\bar{S}\bar{h}^3k^4 \nonumber
	\\& \hskip1.5cm + \bar{J}\dfrac{\Gamma - 1}{\Gamma + |k|\left(\bar{h} + \dfrac{1}{\chi}\right)}
	\left[
		 \epsilon\dfrac{5}{24}Re\sin\beta\bar{E}\bar{h}^4k|k| 
		-i\bar{E}|k|
		+ i\epsilon\left(\mathcal{M} - V_r  \bar{J}\right)\bar{h}^3k^2|k|
	\right].
\end{align}
Notably, the evaporation regime parameter $\Gamma$ appears proportional to the last four terms which in order of appearance in Eq. \eqref{dispersion_relationship_general} represent wave dispersion, mass loss, the diffusion-dependent component of the Marangoni effect, and vapor recoil. 
As in previous literature \cite{burelbach1988nonlinear,joo1991}, the Marangoni effect in the volatile film is coupled to the evaporation process, however, under our formulation it has two distinct components. (i) The first is driven by interfacial temperature gradients caused by the unequal heat flux arriving at the interface from the hot wall due to variations in the film's height and is tied to surface curvature. Therefore, it is proportional to $k^2$.
(ii) On the other hand, the second term is linked to the evaporation regime and depends on the intensity of the diffusion effects and the resulting vapor density gradient above the liquid interface, alongside the interface's curvature. Hence, it is proportional to $k^2|k|$.
Interestingly, unlike previous works which deal exclusively with the transfer-rate-limited evaporation regime \cite{burelbach1988nonlinear,joo1991,mohamed_luca_21}, all of the terms in the dispersion relationship [Eq. \eqref{dispersion_relationship_general}] are proportional to the wavenumber $k$, and are therefore considered to be associated with the instability. 
For our subsequent analyses, we will assign $\epsilon = 0.05$, $\bar{S} = 0.075$, and $\chi = 10$, unless stated otherwise. Additionally, we will assign the value of $\Gamma=1000$ to represent the diffusion-limited-regime since it is sufficiently large to approximate the limit of $\Gamma\rightarrow\infty$.


\section{Temporal stability analysis}\label{sec_temporal_stability}
For a temporal stability analysis, Eq. \eqref{dispersion_relationship_general} admits a real wavenumber $k$ and a complex angular frequency $\omega$, and yields expressions for the temporal growth rate and phase speed.
The expression for the temporal growth rate allows the investigation of the influence of the evaporation regime on the stability of the film and highlights some of its special cases.
Moreover, it reveals the impact of the evaporation regime on the mass loss instability, and, in particular, the effects arising from film thinning.
On the other hand, the expression for the phase speed permits examining the influences of the evaporation regime and film thinning on the phase speed.
\color{black}
\subsection{The impact of the evaporation regime on the stability of the film}
In the temporal analysis, Eq. \eqref{dispersion_relationship_general} takes the form $\omega(k) = \omega_r + i\omega_i$, where $\omega_r=ck$ is the angular frequency written as
\begin{align}\label{omega_r_general}
	\omega_r = \Bigg[Re\sin\beta\bar{h}^2  + \epsilon\dfrac{5}{6}Re\sin\beta\bar{E}\bar{J}\bar{h}^3 + \epsilon\dfrac{5}{24}\dfrac{\Gamma - 1}{\Gamma + |k|\left(\bar{h} + \dfrac{1}{\chi}\right)}Re\sin\beta\bar{E}\bar{J}\bar{h}^4|k|\Bigg]k,
\end{align}
and $\omega_i$ is the temporal growth rate reading
\begin{multline}\label{omega_i_general}
	\omega_i = \epsilon\left[\dfrac{2}{15}\left(Re\sin\beta\right)^2 \bar{h}^6 + \mathcal{M}\bar{J} \bar{h}^2 - \dfrac{1}{3}Re\cos\beta\bar{h}^3 - \bar{S}\bar{h}^3k^2 \right]k^2
	\\+ \bar{J}\dfrac{\Gamma - 1}{\Gamma + |k|\left(\bar{h} + \dfrac{1}{\chi} \right)}
	\Big[
	-\bar{E}|k|
	+ \epsilon\left(\mathcal{M} - V_r \bar{J}\right)\bar{h}^3k^2|k|
	\Big].
\end{multline}
The evaporation regime parameter $\Gamma$ appears proportional to three terms in the temporal growth rate which, in order of appearance in Eq. \eqref{omega_i_general}, represent the contributions of (i) mass loss, (ii) the diffusion-dependent Marangoni effect, and (iii) vapor recoil. Notably, the type of contribution each of these terms serves to the growth rate, whether stabilizing or destabilizing, depends on whether the evaporation regime parameter $\Gamma$ is smaller or larger than 1, as indicated by the change in their algebraic signs in Eq. \eqref{omega_i_general}. Moreover, the last two terms in Eq. \eqref{omega_i_general} do not appear in Sultan \textit{et al.}'s \cite{sultan2005evaporation} derivation.
Note that the temporal growth rate produced by Eq. \eqref{omega_i_general} is validated against a numerically computed counterpart, which is extracted from the numerical simulation of the coupled liquid-vapor system formed by Eqs. \eqref{vapor_system_ND} and Eq. \eqref{Benney_equation}, for further details refer to App. \ref{appendix_numerical}.
\color{black}
When $\Gamma < 1 $, the evaporation process is dictated by thermodynamic disequilibrium at the interface and the roles of the evaporative mass loss and vapor recoil are destabilizing, which is in accordance with what is found in the literature dealing with the transfer-rate-limited regime corresponding to $\Gamma \rightarrow 0$ \cite{burelbach1988nonlinear,joo1991,sultan2005evaporation}. 
However, under this evaporation regime the local evaporation rate across the interface is dictated by the proximity to the hot wall below and is therefore higher at the troughs than at the crests, and hence the local latent cooling is also higher at the troughs. 
This results in a reduction in the temperature-dependent surface tension gradient between the troughs and the crests and ultimately a tempering of the Marangoni instability. This attenuating effect is evident in Eq. \eqref{omega_i_general} in the negative sign of its diffusion-dependent component when $\Gamma < 1$.

On the other hand, when $\Gamma > 1$, the evaporation process is largely dependent on the interface curvature and the associated diffusion of vapor into the ambient gas phase above the liquid interface. We find in this case that the contributions of mass loss and vapor recoil are stabilizing since the vapor flux is higher at the crests than at the troughs due to the sharp edge effect \cite{deegan1997capillary}, i.e., their convex curvature, which matches previous findings \cite{deegan1997capillary,sultan2004diffusion,sultan2005evaporation}. Accordingly, the latent cooling in this regime is higher at the crests rather than the troughs which reduces their local temperature, and thus increases the surface tension gradient across the film. Consequently, the diffusion-dependent component of the Marangoni instability plays a reinforcing role in this evaporation regime and results in a substantially more unstable liquid film. 

\begin{figure}[tp]
	\centering	\includegraphics[width=0.355\textwidth]
	{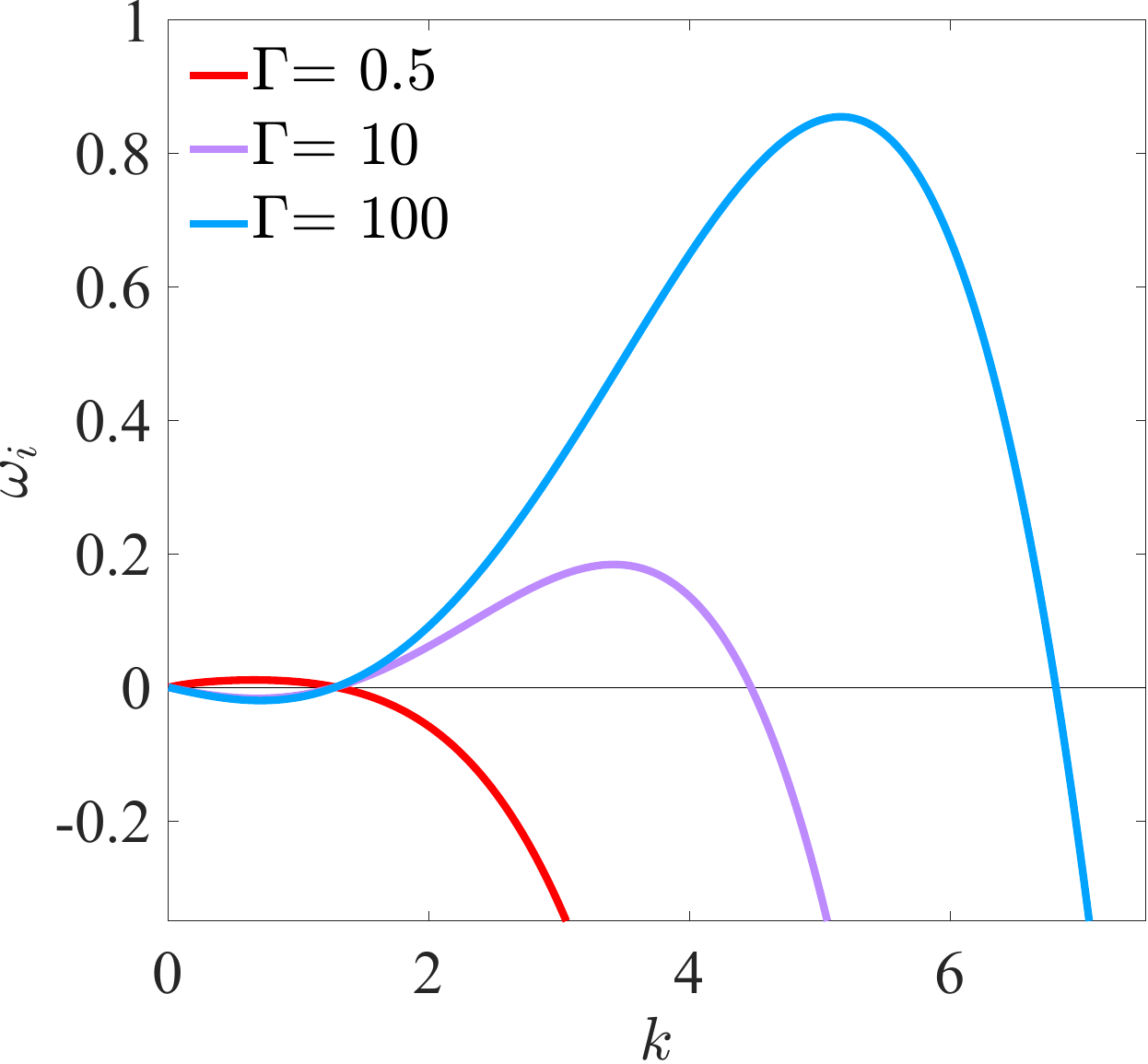}
	\caption{Temporal growth rates for $Re = 2$, $\beta = 25$, $\mathcal{M} = 0.7$, $V_r = 0.1$, $\bar{E}=0.05$,
		and increasing values of $\Gamma$ representing the increasing importance of vapor diffusion.}	\label{evaporation_regime_h1}
\end{figure}

This destabilizing influence of vapor diffusion can be seen in Fig. \ref{evaporation_regime_h1}, where we see that increasing the value of $\Gamma$, which represents the heightening of the influence of vapor diffusion on the evaporation process, results in significant increases in the temporal growth rate $\omega_i$ and the widening of the range of unstable wave numbers. 
Additionally, we find that the Marangoni instability is reinforced by the vapor recoil and mass loss effects when $\Gamma < 1$, as they all contribute positively to the instability in Eq. \eqref{omega_i_general} under this condition, which is in accordance the results of Joo \textit{et al.} \cite{joo1991} and Burelbach \textit{et al.} \cite{burelbach1988nonlinear}.

\begin{figure}[tp]
	\centering
	\begin{minipage}[c]{0.315\textwidth}
		\centering
		\hspace{0.1cm}
		\vspace*{\fill}
		\includegraphics[width=0.95\textwidth]{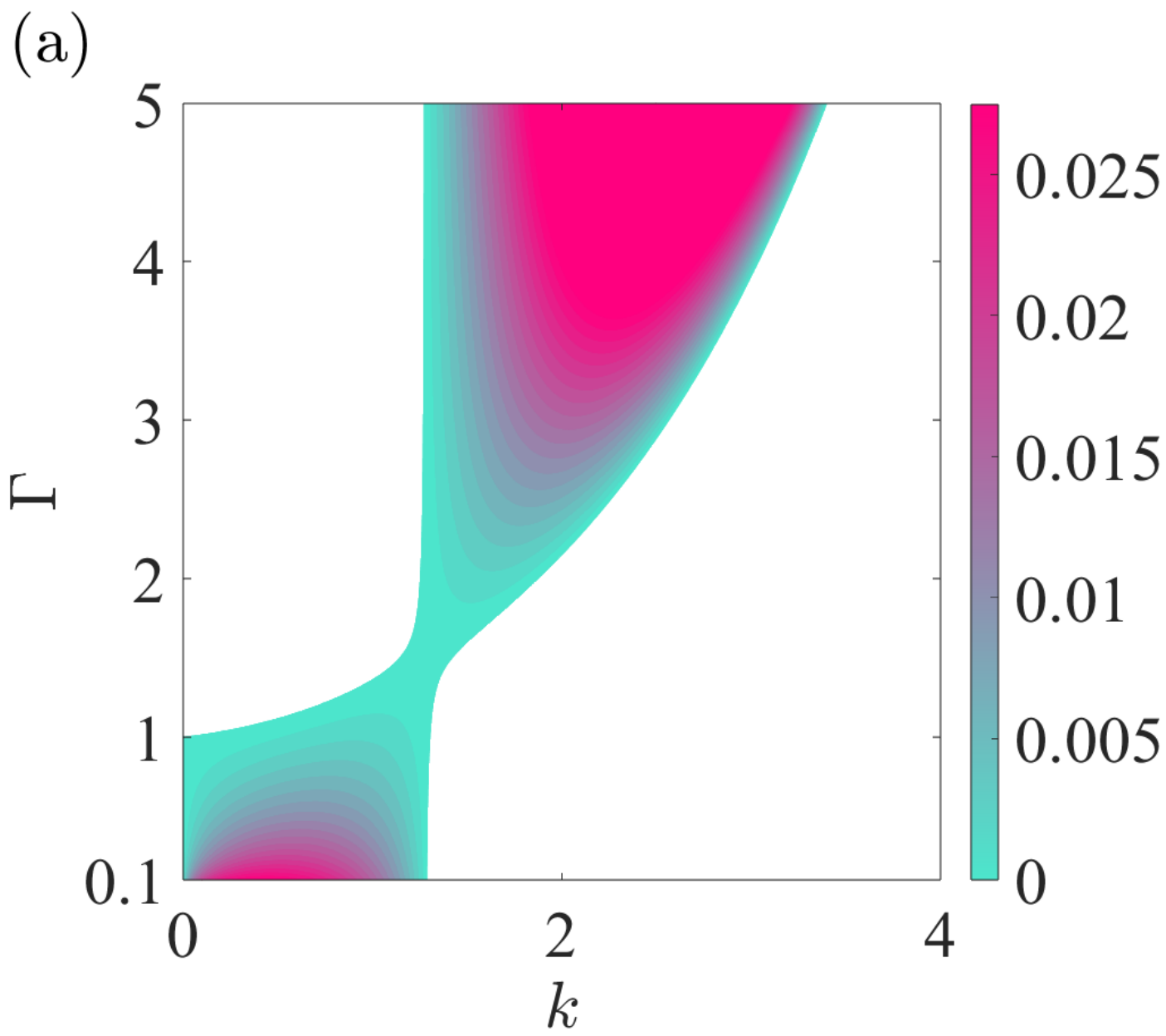}
		\vspace*{\fill}
	\end{minipage}
	\hspace{0.05\textwidth}
	\begin{minipage}[c]{0.615\textwidth}
		\centering
		\vspace*{\fill}
		\begin{subfigure}[c]{0.49\textwidth}
			\centering
			\includegraphics[width=\textwidth]{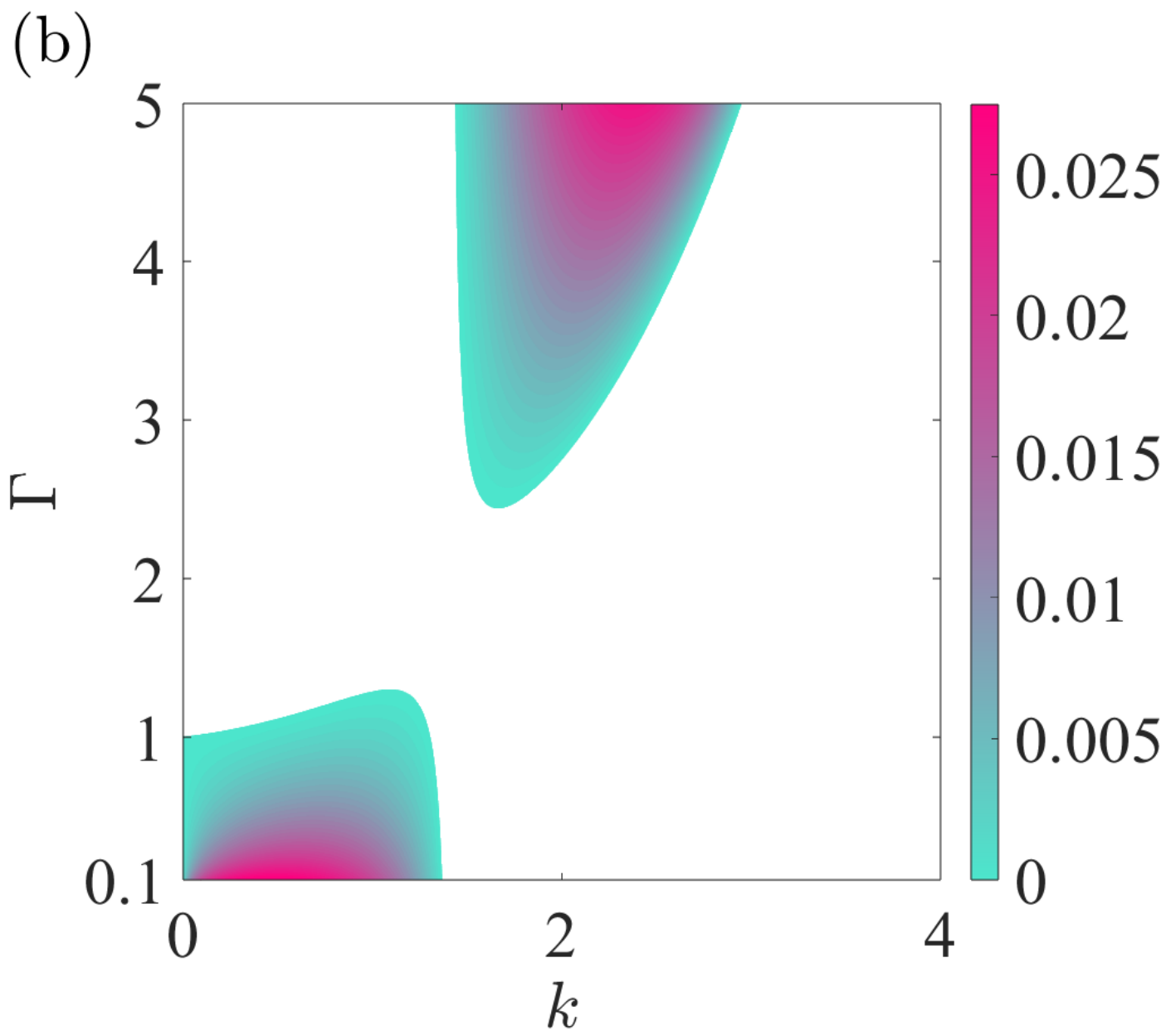}
		\end{subfigure}
		\hfill
		\begin{subfigure}[c]{0.49\textwidth}
			\centering
			\includegraphics[width=\textwidth]{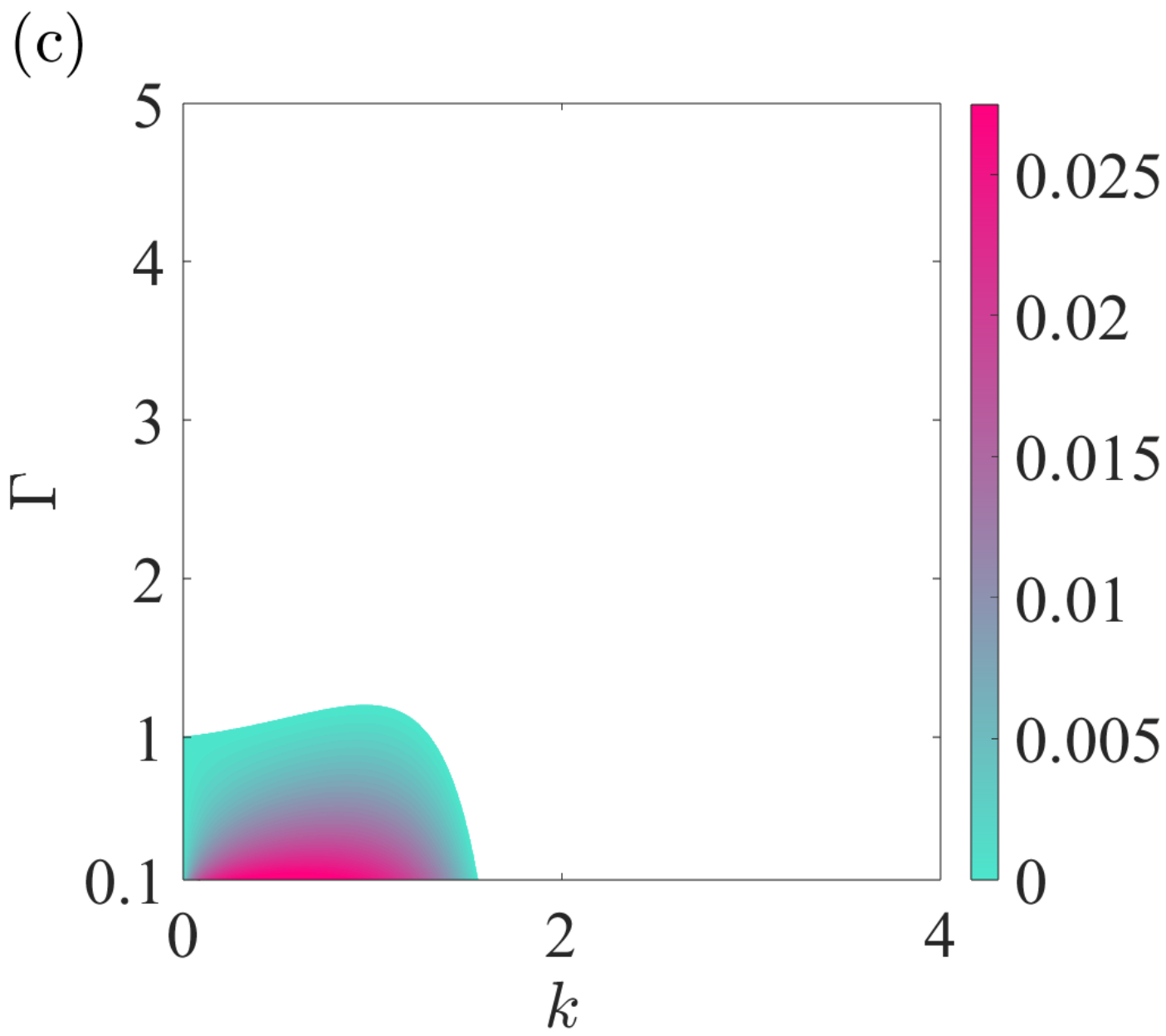}
		\end{subfigure}
		\\
		\begin{subfigure}[c]{0.49\textwidth}
			\centering
			\includegraphics[width=\textwidth]{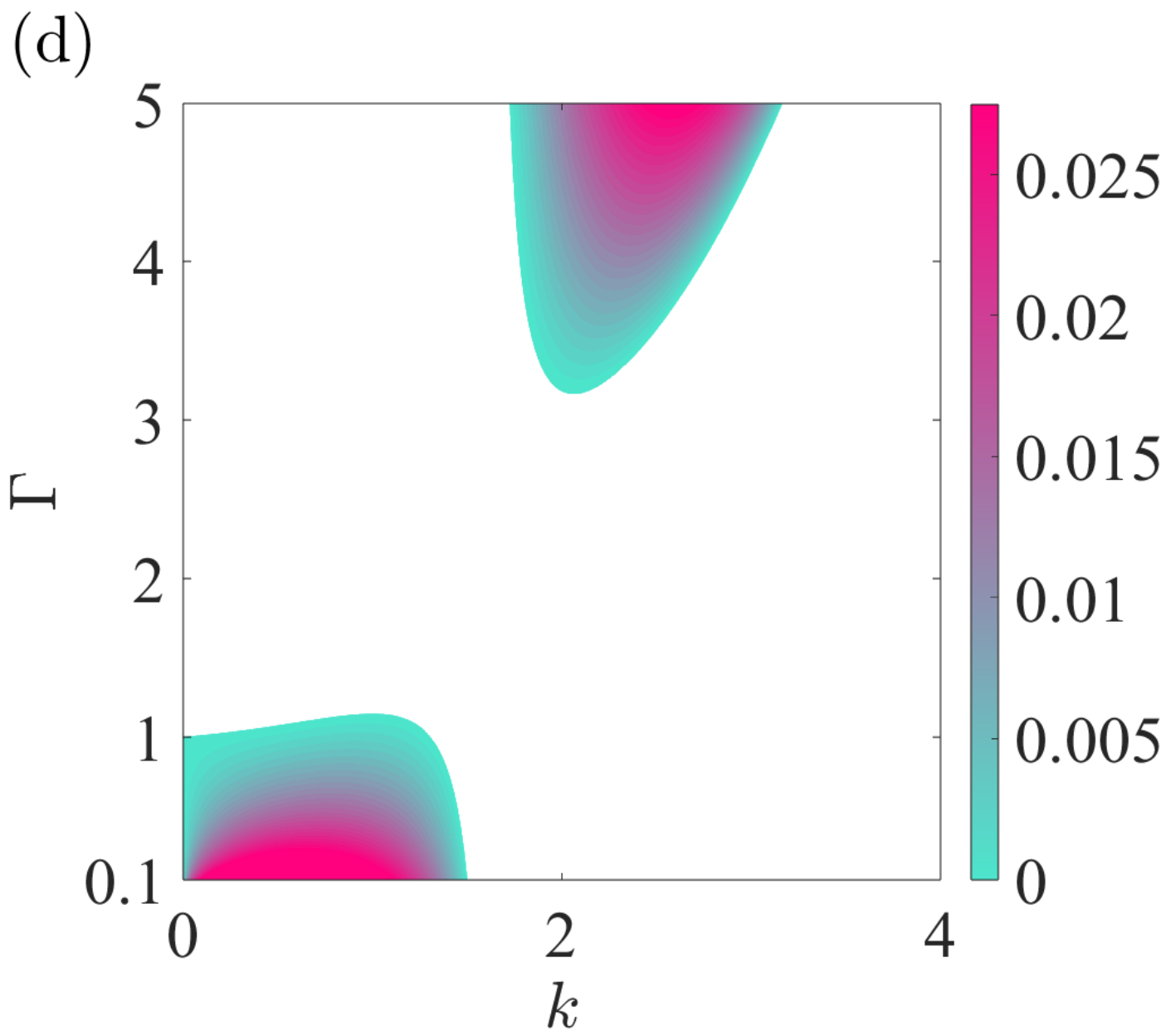}
		\end{subfigure}
		\hfill
		\begin{subfigure}[c]{0.49\textwidth}
			\centering
			\includegraphics[width=\textwidth]{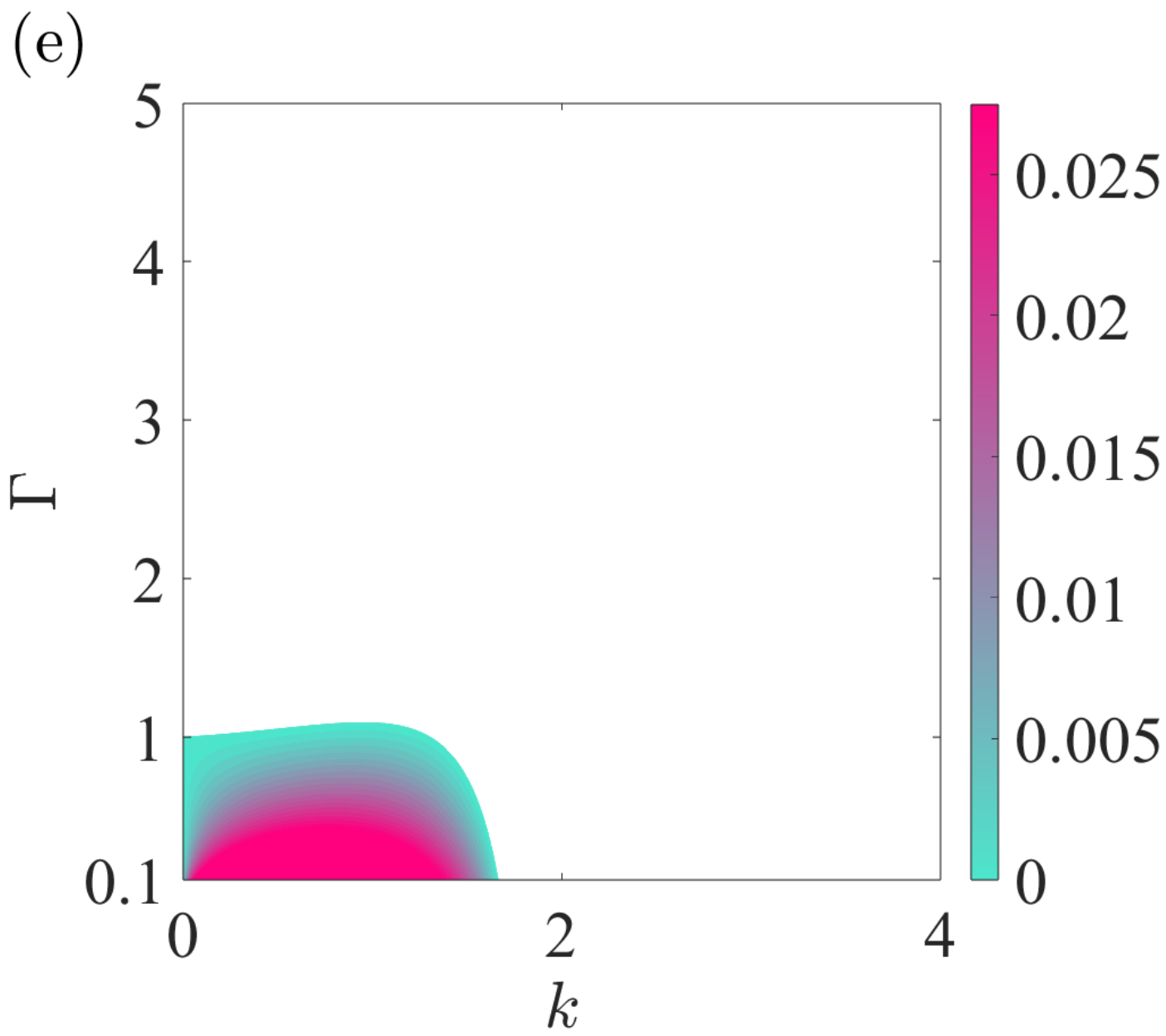}
		\end{subfigure}
		\vspace*{\fill}
	\end{minipage}
	\caption{
		Temporal growth rates in the $(k,\Gamma)$ space as influenced by the strength of evaporation effects for $Re = 2$, $\beta = 25$, and $\mathcal{M} = 0.7$. 
		(a) Reference case with relatively weak evaporation ($V_r = 0.1$, $\bar{E} = 0.05$).
		Figs. (b) and (c) depict the effect of stronger vapor recoil, $V_r = 0.2$ and $V_r = 0.4$, respectively.
		Figs. (d) and (e) depict the effect of stronger mass loss, $\bar{E} = 0.075$ and $\bar{E} = 0.1$, respectively.
	}
	\label{Marangoni_vs_Evaporation}
\end{figure}

Conversely, when the evaporation process is dominated by diffusion $(\Gamma > 1)$ the Marangoni effect has a strong destabilizing influence that is suppressed by vapor recoil and evaporative mass loss, as shown in Fig. \ref{Marangoni_vs_Evaporation}. 
When evaporation effects are relatively weak $(V_r = 0.1, \bar{E} = 0.05)$ as in Fig. \ref{Marangoni_vs_Evaporation}(a), we observe two regions of intense instability in the $(k,\Gamma)$ space: a region of instability for small values of $\Gamma$ which is caused by the combination of the subdued Marangoni instability, mass loss, and vapor recoil, as well as a significantly stronger region of instability at relatively larger values of $\Gamma$, which arises mostly from the Marangoni effect as amplified by latent cooling. 
In Fig. \ref{Marangoni_vs_Evaporation}(b) vapor recoil is stronger $(V_r = 0.2)$ and has two opposite effects depending on the evaporation regime. At small values of $\Gamma$, the stronger vapor recoil strengthens the instability, whereas at relatively larger values of $\Gamma$ it weakens it. These two opposing trends continue as the value of $V_r$ is further increased to $V_r = 0.4$, as seen in Fig. \ref{Marangoni_vs_Evaporation}(c), where we observe further intensification of the instability in the kinetically-dominated regime, alongside a complete suppression of the perturbation in the diffusion-dominated evaporation regime.
Likewise, the effect of evaporative mass loss on the $(k,\Gamma)$ parameter space is shown in Figs. \ref{Marangoni_vs_Evaporation}(d) and \ref{Marangoni_vs_Evaporation}(e), where again the contrast in its role and the dependence of its influence on the evaporation regime is evident. 
Figs. \ref{Marangoni_vs_Evaporation}(d) and \ref{Marangoni_vs_Evaporation}(e) show the gradual stabilization (destabilization) that results from raising the value of $\bar{E}$ when $\Gamma$ is relatively large (small).

\subsection{Special cases in the evaporation regime}
The evaporation regime contains three unique situations represented by three special cases of  the temporal growth rate [Eq. \eqref{omega_i_general}].
Firstly, the case of $\Gamma \rightarrow 0$ represents the transfer-rate-limited regime where the liquid's vapor does not appear as a variable in the system and the role of vapor diffusion is excluded. This is reflected in the elimination of $|k|$ from Eq. \eqref{omega_i_general} since it is associated with the vapor gradient above the liquid interface, and hence in the transfer-rate-limited regime Eq. \eqref{omega_i_general} reads 
\begin{align}\label{omega_i_TL}
	\omega_i = \dfrac{\bar{E}}{\Big(\bar{h} + \dfrac{1}{\chi}\Big)^2}
	+ \epsilon k^2\Bigg[
	\dfrac{2}{15}\left(Re\sin\beta\right)^2 \bar{h}^6 
	+ \dfrac{\mathcal{M}}{\chi}\dfrac{\bar{h}^2}{\Big(\bar{h} + \dfrac{1}{\chi}\Big)^2} 
	+ \dfrac{V_r\bar{h}^3}{\Big(\bar{h} + \dfrac{1}{\chi}\Big)^3}
	- \dfrac{Re}{3}\cos\beta\bar{h}^3  - \bar{S}\bar{h}^3k^2 
	\Bigg],
\end{align}
in which we have substituted explicitly for the base vapor flux $\bar{J}$. Equation \eqref{omega_i_TL} recovers the exact expression for the temporal growth rate in Joo \textit{et al.} \cite{joo1991} with the exception of an extraneous higher-order term, where $1/\chi$ has taken the place of their interfacial nonequilibrium parameter.
Remarkably, this limit naturally allows unifying the two terms representing the Marangoni instability in Eq. \eqref{omega_i_general} into a single term, which is compatible with the singular dependence of both drivers of the surface tension gradient on the film thickness, namely the heat flux across the film and the vapor flux from the interface.
Note that when the limit of $\Gamma \rightarrow 0$ is taken in the original work of Sultan \textit{et al.} \cite{sultan2005evaporation}, only the special case of quasi-equilibrium conditions and $\bar{h}= 1$  is recovered from Burelbach \textit{et al.}'s \cite{burelbach1988nonlinear} work.

Secondly, as in the work of Sultan \textit{et al.} \cite{sultan2005evaporation}, the case of $\Gamma \rightarrow \infty$ represents the diffusion-limited regime where the vapor gradient above the film is a primary influence on its stability, which is manifested in the expression for the temporal growth rate under this limit which now reads: 
\begin{align}\label{omega_i_diff}
	\omega_i = -\bar{E}\bar{J}|k|
	+ \epsilon k^2\Bigg[
	\dfrac{2}{15}\left(Re\sin\beta\right)^2 \bar{h}^6 
	+ \mathcal{M}\bar{J}\left(\bar{h}^2 + \bar{h}^3|k|\right)
	- V_r\bar{J}^2\bar{h}^3|k|
	- \dfrac{Re}{3}\cos\beta\bar{h}^3  - \bar{S}\bar{h}^3k^2 
	\Bigg].
\end{align}
Note that the term representing the Marangoni instability is maximized in under these conditions since the coefficient associated with its diffusion-dependent component takes its peak value when $\Gamma \rightarrow \infty$, which is consistent with the fact that the augmenting role of latent cooling is maximized when the evaporation process is entirely governed by diffusion.

\begin{figure}[tp]
	\centering
	\includegraphics[width=0.55\linewidth]{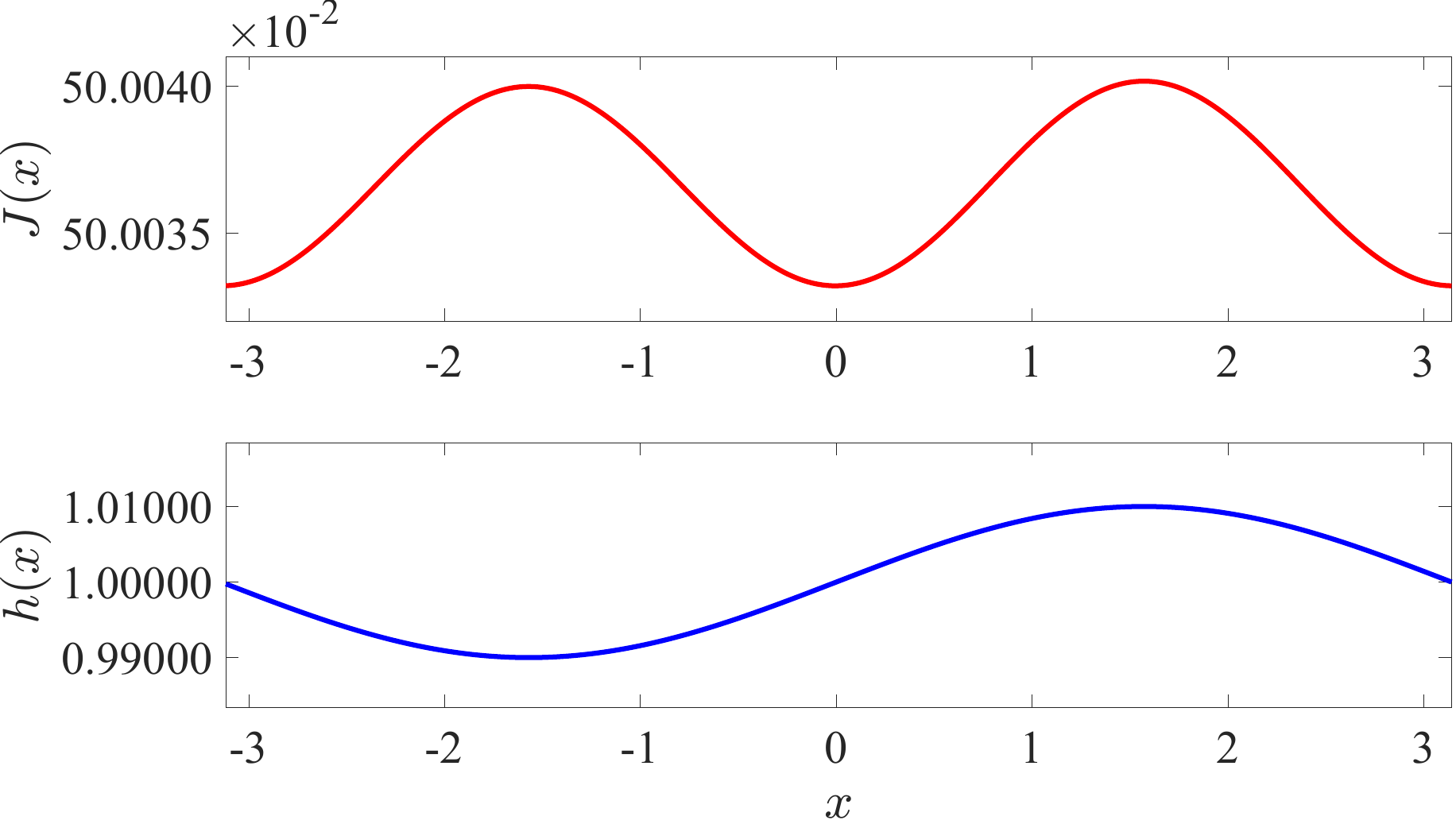}
	\caption{Plot of the perturbed liquid interface (bottom) and the corresponding numerically computed vapor flux (top) for 
		$\Gamma = 1$, $k = 1$, and a grid size of $(N_x,N_y) = (256,64)$.}
	\label{Gamma1_flux}
\end{figure}

Lastly, when $\Gamma=1$, the last three terms in Eq. \eqref{omega_i_general} vanish and it reduces to
\begin{align}\label{omega_i_gamma_one}
	\omega_i = \epsilon\left[\dfrac{2}{15}\left(Re\sin\beta\right)^2 \bar{h}^6 - \dfrac{Re}{3}\cos\beta\bar{h}^3 + \mathcal{M}\bar{J} \bar{h}^2 - \bar{S}\bar{h}^3k^2 \right]k^2,
\end{align}
which represents a volatile film where the increased evaporation rate at the troughs due to their increased proximity to the hot wall is precisely balanced by the heightened evaporation rate at the crests due to the higher diffusion rate caused by their convex curvature. Interestingly, this special case represents an evaporating film that is naturally devoid of the vapor recoil and mass loss instabilities. As for the Marangoni instability, since the evaporative latent cooling is equalized across the troughs and the crests, it neither diminished nor augmented and is therefore intermediate in strength between the transfer-rate-limited and the diffusion-limited cases.
We verify that the vapor mass flux is indeed equal at the crests and the troughs of the film when $\Gamma = 1$ by numerically solving the vapor system [Eqs. \eqref{vapor_system_ND}] for a perturbed interface described by $h(x) = 1 + 0.01\sin(kx)$, as shown in Fig. \ref{Gamma1_flux}. Due to this particular property, we will refer to this case as the \textit{balanced flux} regime.

\subsection{The impact of the evaporation regime on the mass loss instability}
The role that evaporative mass loss has on the temporal stability of the film is tied to the evaporation regime through several facets. In the complete absence of vapor diffusion effects i.e., the transfer-rate-limited regime, the evaporation process is kinetically driven and, thus, the film's troughs which are closer to the hot wall will decay faster than the crests, since they experience relatively higher evaporation rates. This results in perturbation growth and ultimately a destabilizing influence for evaporative mass loss, as reported in Refs. \cite{burelbach1988nonlinear,joo1991}. Moreover, since the evaporation process here is one-sided, i.e., does not account for a gas phase above the liquid film, there is no vapor gradient restricting the departure of molecules from the interface, and thus, their ability to depart the interface is equal across the film regardless of the local interface curvature, as it is only contingent on the local film height.

These two aspects of the mass loss effect are reflected in the expression for the temporal growth representing this limit [Eq. \eqref{omega_i_TL}], where we find, similar to previous works dealing with the transfer-rate-limited evaporation regime \cite{burelbach1988nonlinear,joo1991,sultan2005evaporation, mohamed_luca_21}, that the mass loss term (i) contributes positively to the temporal growth rate, and (ii) is independent of the wave number $k$, which is consistent with the effect being independent of the interface's curvature. Therefore, this term is typically neglected when computing the temporal growth rate \cite{joo1991,mohamed_luca_21} in this regime since it is not directly related to the perturbation, and its contribution is restricted to an algebraic shift in the temporal growth rate across all wave numbers. Thus, we keep in accordance with this convention and set $\bar{E}$ to zero when investigating the stability of the film in the transfer-rate-limited regime.

On the other hand, accounting for the liquid's vapor introduces a vapor gradient above the liquid film which varies with its curvature, and hence, the effect of mass loss is not realized equally across the interface since the propensity of the  molecules to depart from it depends on the variation in this gradient. Thus, in the presence of vapor diffusion, the role of mass loss is properly tied to the instability through the film height perturbation.
Again, these properties are captured in the associated expressions for the temporal growth rate [Eqs. \eqref{omega_i_general} and \eqref{omega_i_diff}], where the mass loss term is proportional to the wave number. Consequently, we retain this term when $\Gamma > 0$.

Ultimately, the influence that mass loss has in the two-phase system, whether stabilizing or destabilizing, is determined by the relative strengths of the kinetic and diffusion phenomena. The effect is destabilizing when kinetic effects dominate $(\Gamma < 1)$, as the film's troughs will overall evaporate more intensely due to their proximity to the hot wall, despite the hindrance posed by the unfavorable vapor gradient. On the contrary, the effect is stabilizing when vapor diffusion effects prevail $(\Gamma > 1)$, as the film's crests will evaporate more intensely due to their convex curvature, despite their adjacency to the hot wall.

\begin{figure}[tp]
	\centering
	\begin{subfigure}[h]{0.355\textwidth}
		\includegraphics[width=\textwidth]
		{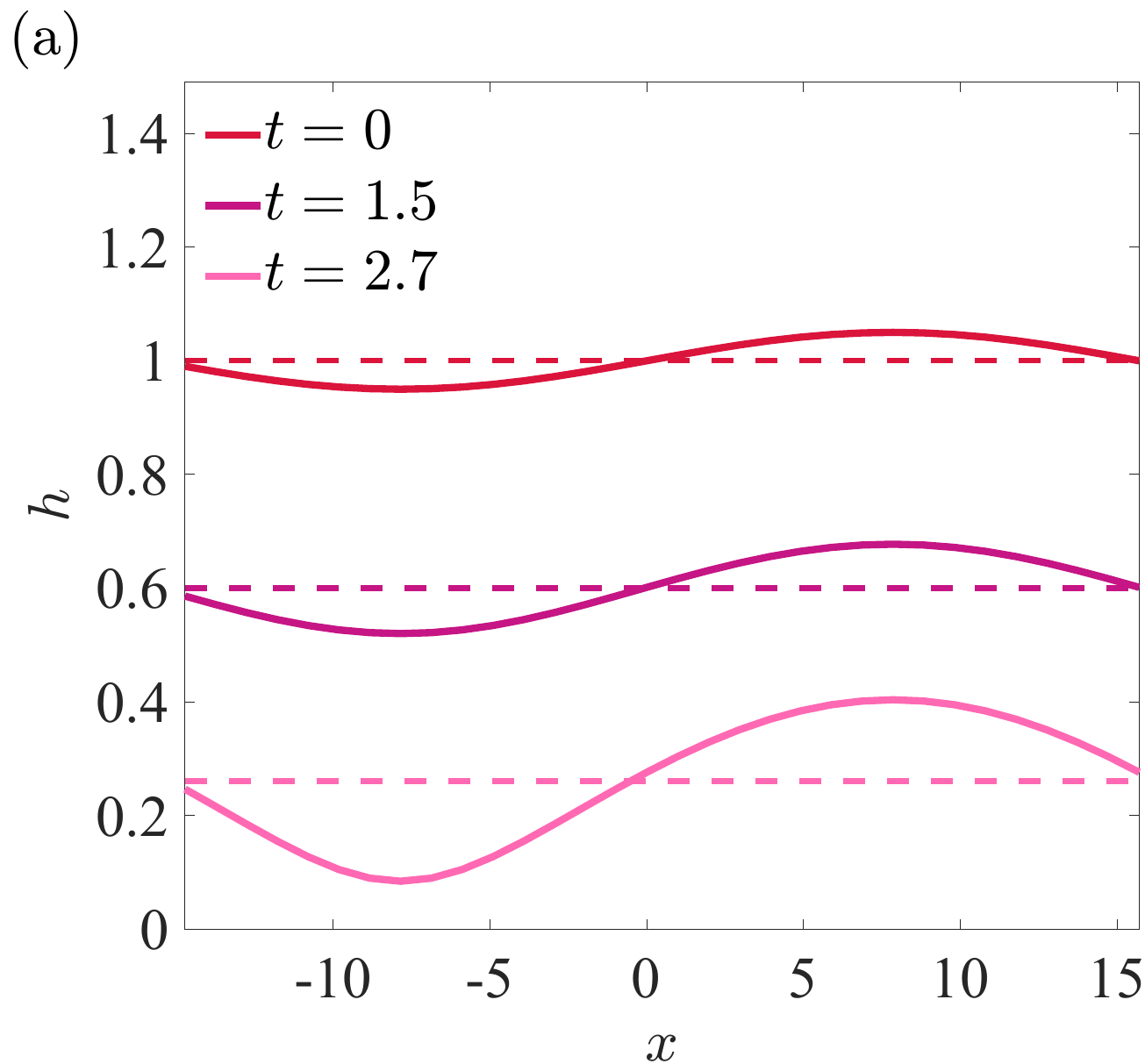}
	\end{subfigure}
	\hskip 1.5cm
	\begin{subfigure}[h]{0.355\textwidth}
		\includegraphics[width=\textwidth]
        {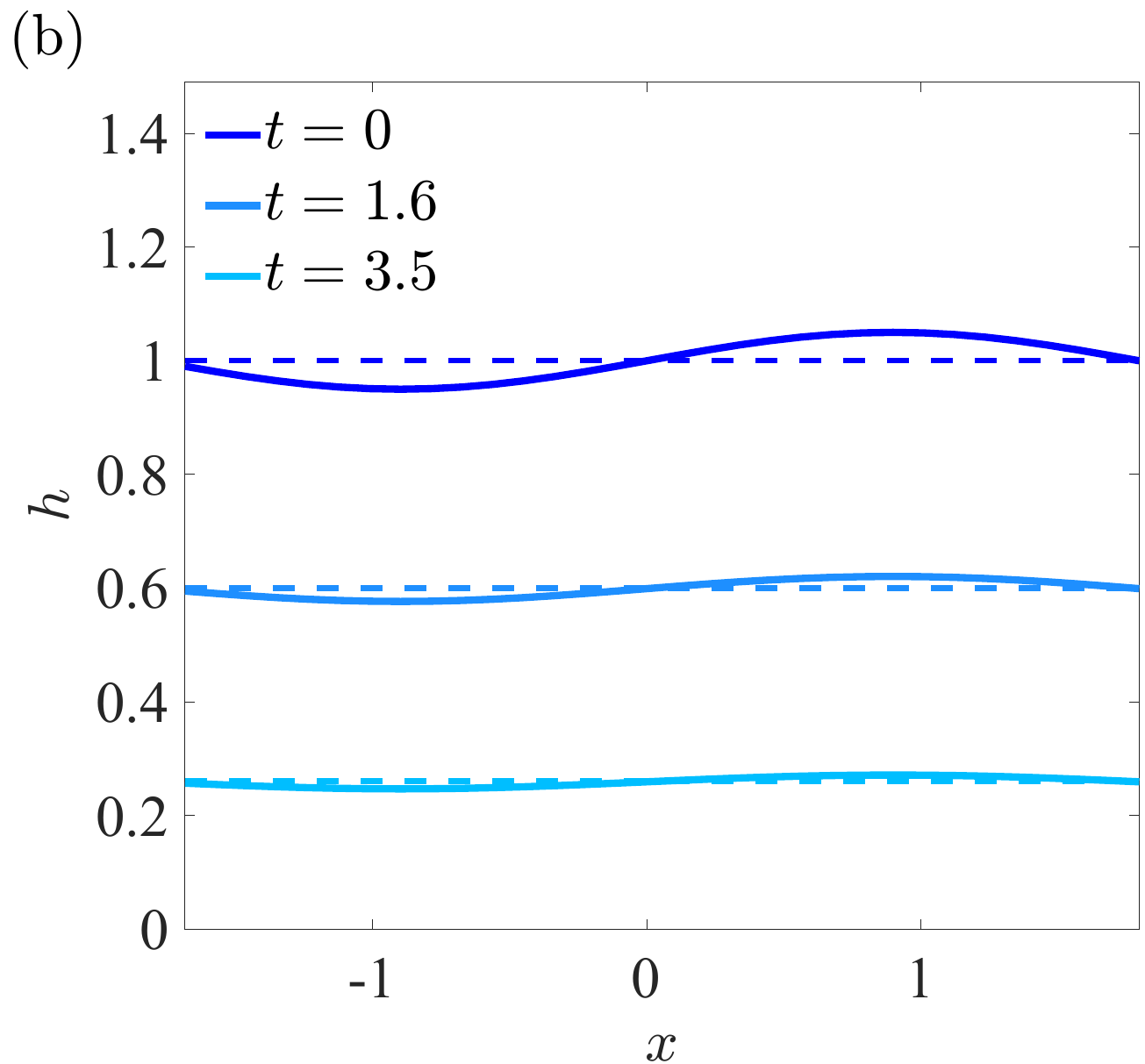}
	\end{subfigure}
	\caption{Numerical simulations of a thinning evaporating film comparing of the role of evaporative mass loss at the two limits of the evaporation regime, for
		$Re = 2$, $\beta = 0$, $\mathcal{M} = 0$, $V_r = 0$, $\bar{E}=0.1$.
		(a) Transfer-rate-limited evaporation, $\Gamma = 10^{-6}$, $k = 0.2$, $(N_x,N_y) = (32, 8)$.
		(b) Diffusion-limited evaporation, $\Gamma = 1000$, $k = 1.75$, $(N_x,N_y) = (32, 1024)$.
		In both cases the simulation started from an initial interface $h = 1 + 0.05\sin(kx)$.
		The dashed lines represent the reducing flat film height at the corresponding time.
	}
	\label{mass_loss_comparison}
\end{figure}

We demonstrate this contrasting effect of evaporative mass loss in the presence of vapor diffusion by simulating the coupled liquid-vapor system through the numerical solution of Eqs. \eqref{vapor_system_ND} and Eq. \eqref{Benney_equation}, where we isolate the effect of evaporative mass loss by neglecting vapor recoil and Marangoni phenomena $(V_r = \mathcal{M} = 0)$, and keeping $\bar{E}\neq 0$. The results of this simulation are presented in Fig. \ref{mass_loss_comparison} where we chose values of $\Gamma$ representing the extremities of the evaporation regime to maximize the observable effect. Fig. \ref{mass_loss_comparison}(a) depicts an evaporating thinning film at different points in time under transfer-rate-limited conditions $(\Gamma \rightarrow 0)$, where the perturbation is observed to grow in time as a result of unequal mass loss across the film. Conversely, Fig. \ref{mass_loss_comparison}(b) depicts the film in the thinning in the diffusion-limited regime $(\Gamma = 1000)$, where the evaporative mass loss is results in the perturbation's decay and,  eventually, the restoration of the flat film.
\subsection{The impact of the evaporation regime on the effects of film thinning}
The effect of the reduction in the base flow height $\bar{h}$ due to evaporation on the thermal instabilities was studied previously by Mohamed \textit{et al.} \cite{mohamed_luca_21}. However, their analysis was restricted to the transfer-rate-limited regime corresponding to $\Gamma \rightarrow 0$. In our work, we also investigate the effect of film thinning in the presence of vapor diffusion effects. The competition between the different physical phenomena influencing the stability of the film as it thins is best understood in the limiting cases of transfer-rate-limited evaporation and diffusion-limited evaporation. 

\begin{figure}[tp]
	\centering
	\begin{subfigure}[t]{0.355\textwidth}
		\includegraphics[width=\textwidth]
		{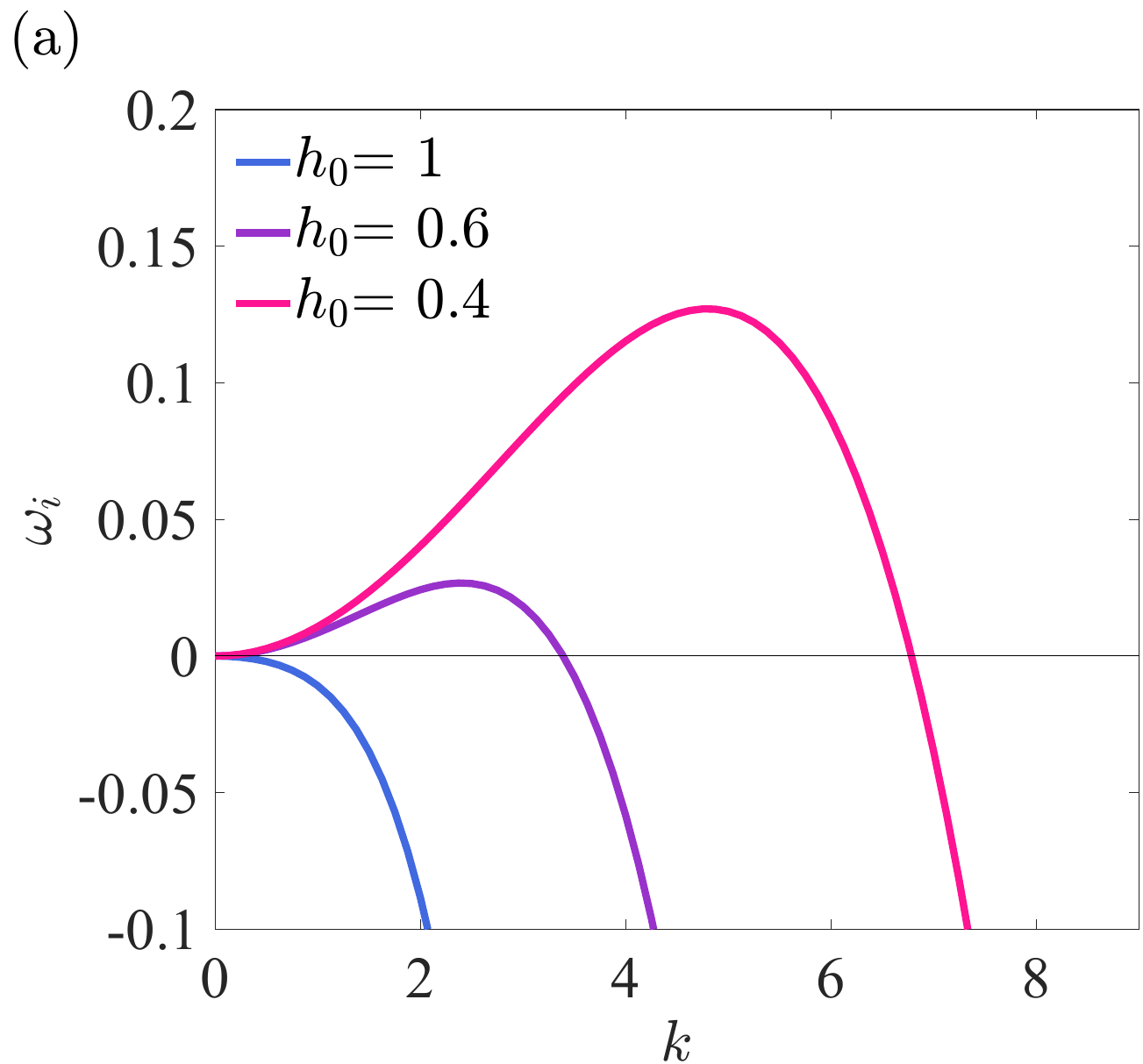}
	\end{subfigure}
	\hskip 1.5cm
	\begin{subfigure}[t]{0.355\textwidth}
		\includegraphics[width=\textwidth]
		{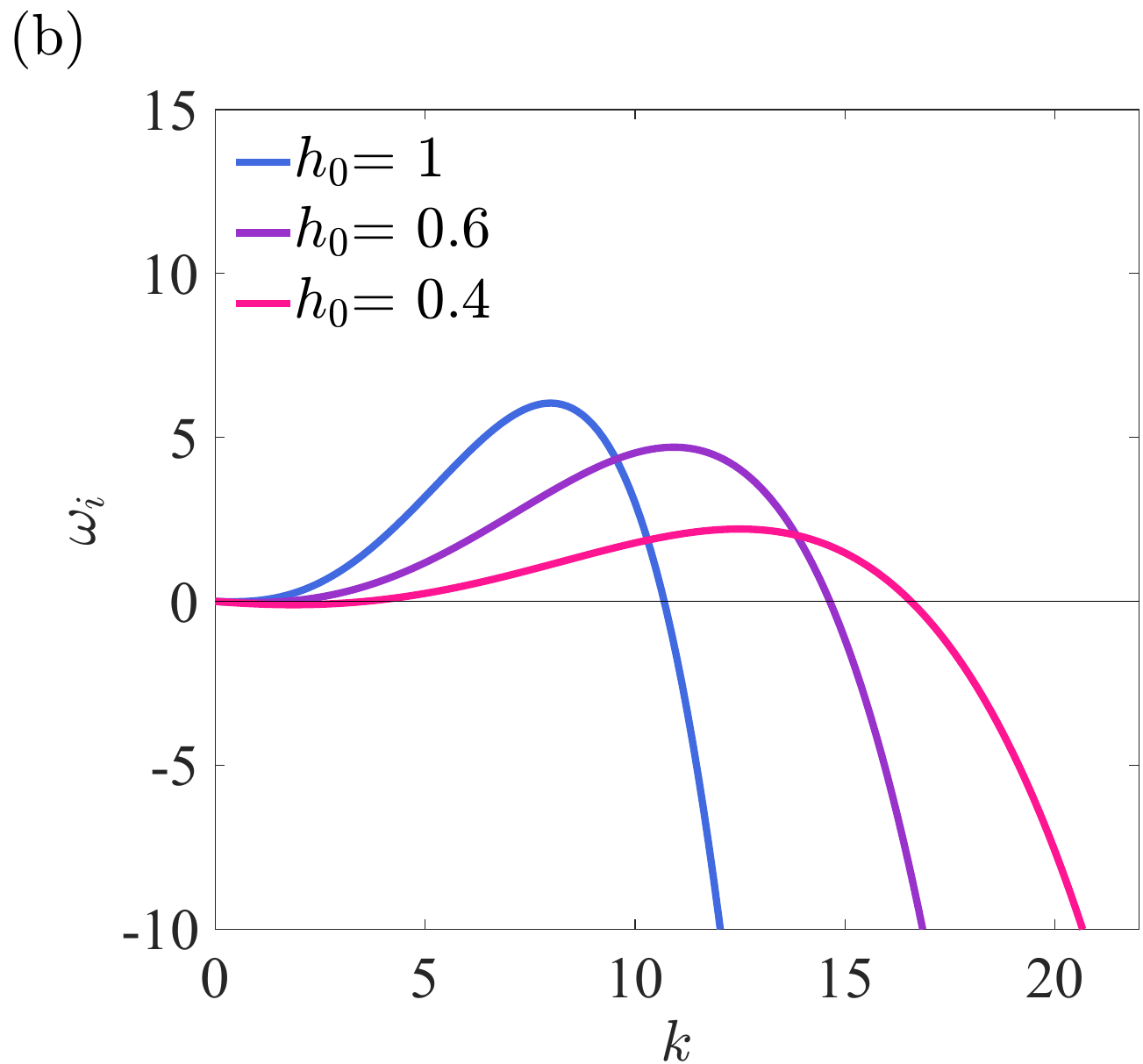}
	\end{subfigure}
	\\
	\begin{subfigure}[t]{0.355\textwidth}
		\includegraphics[width=\textwidth]
		{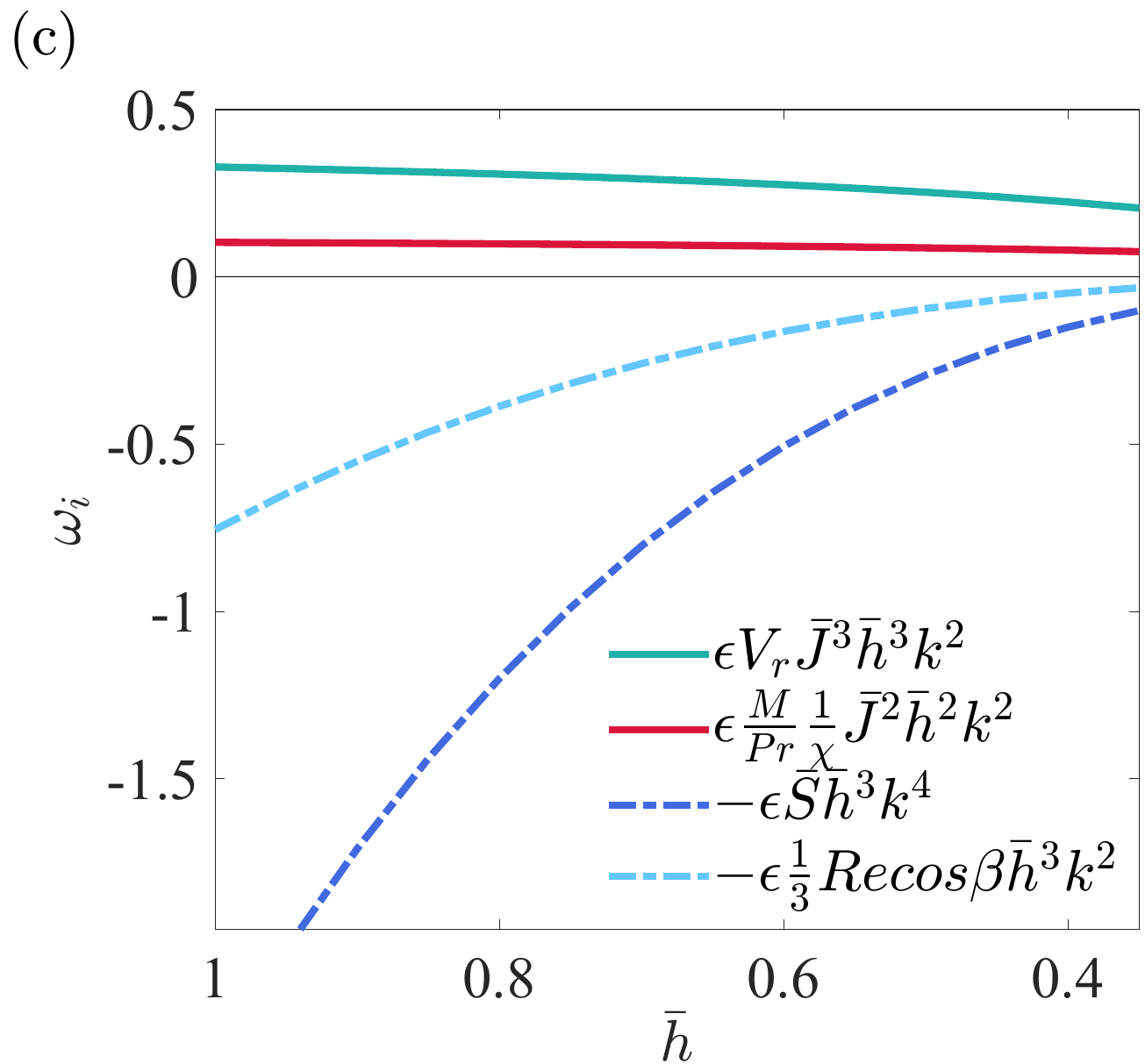}
	\end{subfigure}
	\hskip 1.5cm\label{key}
	\begin{subfigure}[t]{0.355\textwidth}
		\includegraphics[width=\textwidth]
		{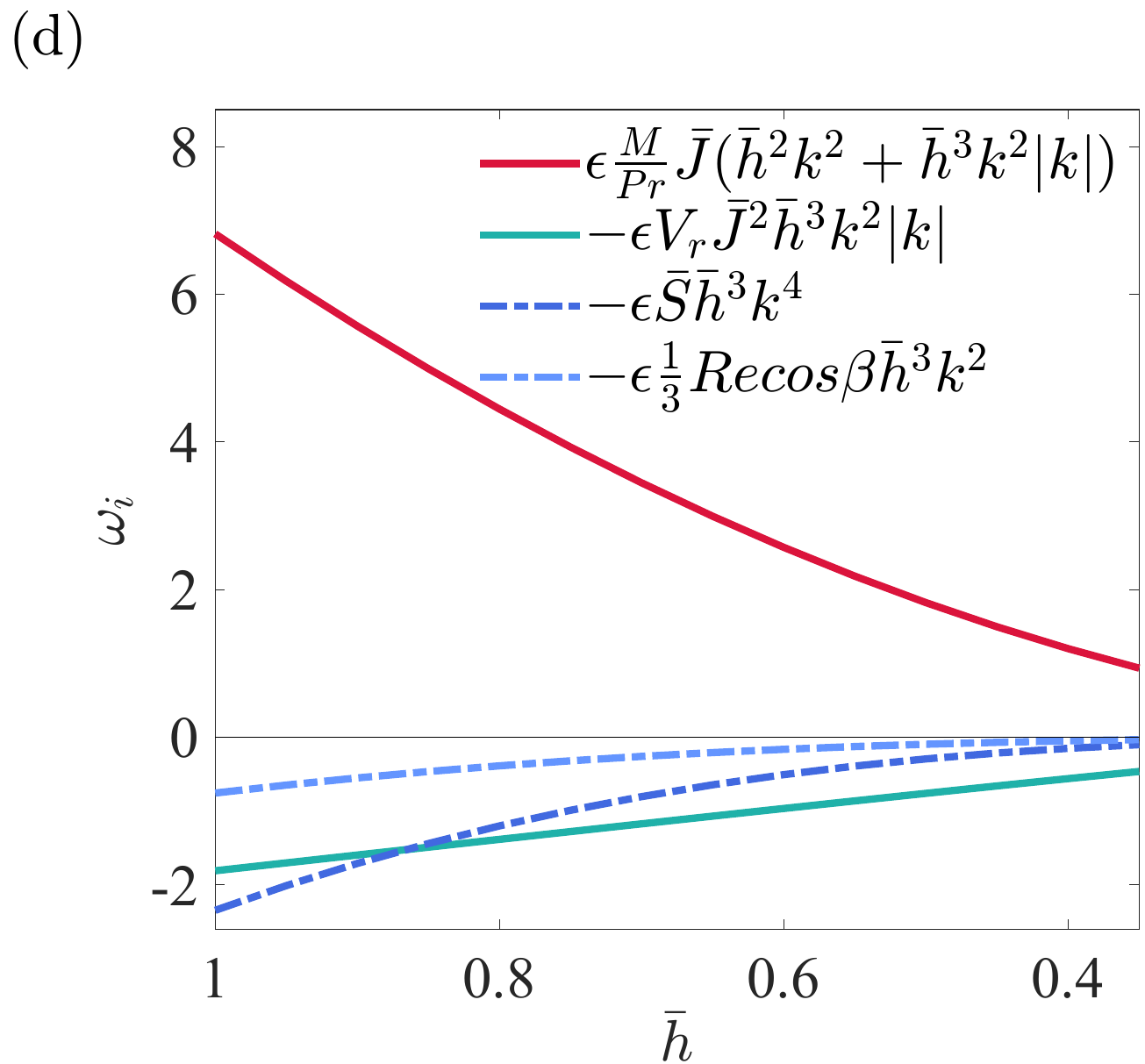}
	\end{subfigure}
	\caption{Comparison of the effect of film thinning on the temporal stability of the film at the two limits of the evaporation regime for 
		$Re = 2$, $\beta = 25$, $\mathcal{M} = 1.25$, and $V_r = 0.35$.
		(a) Temporal growth rates in the transfer-rate-limited regime $(\Gamma = 0$, $\bar{E} = 0)$.
		(b) Temporal growth rates in the diffusion-limited regime $(\Gamma = 1000$, $\bar{E} = 0.05)$.
		(c) Competition between the physical effects in the transfer-rate-limited regime $(\Gamma = 0$, $\bar{E} = 0$, $k = 5)$.
		(d) Competition between the physical effects in the diffusion-limited regime $(\Gamma = 1000$, $\bar{E} = 0.05$, $k = 5)$.
	}
	\label{film_thinning_growthrate}
\end{figure}

The effect of film thinning on the temporal growth rate in the transfer-rate-limited regime $(\Gamma \rightarrow 0)$ is shown in Fig. \ref{film_thinning_growthrate}(a) where we a observe a drastic increase in both the temporal growth rate and range of unstable wave numbers as the film thickness is reduced. This results in the stable film corresponding to $\bar{h} = 1$ becoming increasingly unstable as the value of $\bar{h}$ is decreased, repeating the findings of Ref. \cite{mohamed_luca_21}.
On the other hand, Fig. \ref{film_thinning_growthrate}(b) depicts the temporal growth rate in the case of $\Gamma = 1000$ which represents the diffusion-limited regime, where we observe markedly different behavior as the reduction in $\bar{h}$ causes a reduction in the temporal growth rate indicating a stabilizing effect. Note this is also accompanied by an expansion of the set of unstable wave numbers.
To understand this contrast, we inspect the change in the dominant components of the temporal growth rate with $\bar{h}$ close to the most unstable wave number (denoted $k_{max}$) in the two disparate regimes, as shown in Figs. \ref{film_thinning_growthrate}(c) and (d). We do not include the inertial component in our analysis due to its diminished influence on the thermal instability, or the mass loss component since it is only first order in the wave number while the others are of higher orders. 
Fig. \ref{film_thinning_growthrate}(c) corresponds to Fig. \ref{film_thinning_growthrate}(a) where we plot the two components of the thermal instability in the transfer-rate-limited regime: the Marangoni effect and vapor recoil, along with the two stabilizing phenomena, hydrostatic pressure and surface tension against the base flow height $\bar{h}$. We find that as the film thins the reduction in the stabilizing effects is significantly larger than the decline in the destabilizing ones, yielding an overall destabilization of the film. This is consistent with the order of these terms in the expression for the temporal growth rate corresponding to this regime [Eq. \eqref{omega_i_TL}], as it can be seen that the Marangoni term and vapor recoil terms diminish with $\bar{h}$ in $\sim \mathcal{O}(1)$, whereas the hydrostatic pressure and surface tension terms vary with it in $\sim \mathcal{O}(\bar{h}^3)$. 
Likewise, Fig. \ref{film_thinning_growthrate}(d) corresponds to Fig. \ref{film_thinning_growthrate}(b), where the augmented Marangoni effect remains the sole driver of the thermal instability since vapor recoil becomes a stabilizing factor in the diffusion-limited regime. We notice in this case that the destabilizing Marangoni effect reduces rather fast with $\bar{h}$ in a manner that is comparable to the stabilizing factors it competes against, which now also include vapor recoil, resulting ultimately in an overall  stabilization of the film. 
This behavior again conforms to the order of the terms in the corresponding Eq. \eqref{omega_i_diff}, where the augmented Marangoni term reduces with $\sim \mathcal{O}(\bar{h}^2)$ which is closer to the order of the stabilizing terms with respect to $\bar{h}$. 

This heightened sensitivity of the Marangoni term to the base flow height arises in large part due to the reliance of its diffusion-dependent component on the intensity of latent cooling, which loses its efficacy as the film thins in this evaporation regime. This is because the thinner film corresponds to higher interfacial temperatures, yet the evaporation process and hence the latent cooling, remain constrained by the diffusion rate. This is in stark contrast to the transfer-rate-limited case, where the higher interfacial temperatures correspond immediately to higher evaporation rates.
Notably, the type of stabilization observed in Fig. \ref{film_thinning_growthrate}(b) occurs when the stabilizing evaporation effects, namely vapor recoil and mass loss, are strong enough to compensate for the fast reduction in the hydrostatic pressure and surface tension terms as $\bar{h}$ reduces. 

\begin{figure}[tp]
	\centering
	\begin{subfigure}[t]{0.3\textwidth}
		\includegraphics[width=\textwidth]
		{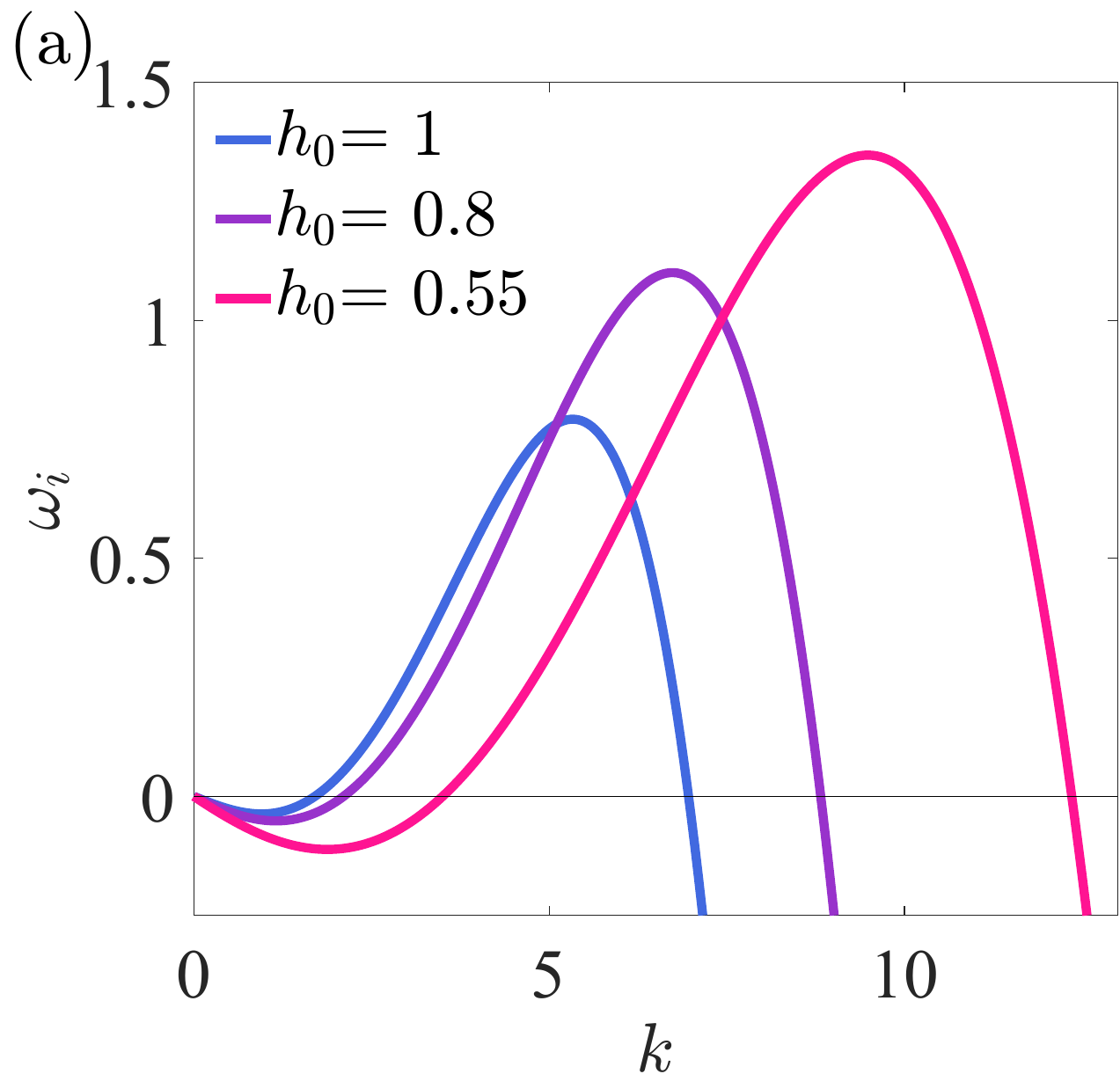}
	\end{subfigure}
	\hskip0.3cm
	\begin{subfigure}[t]{0.3\textwidth}
		\includegraphics[width=\textwidth]
		{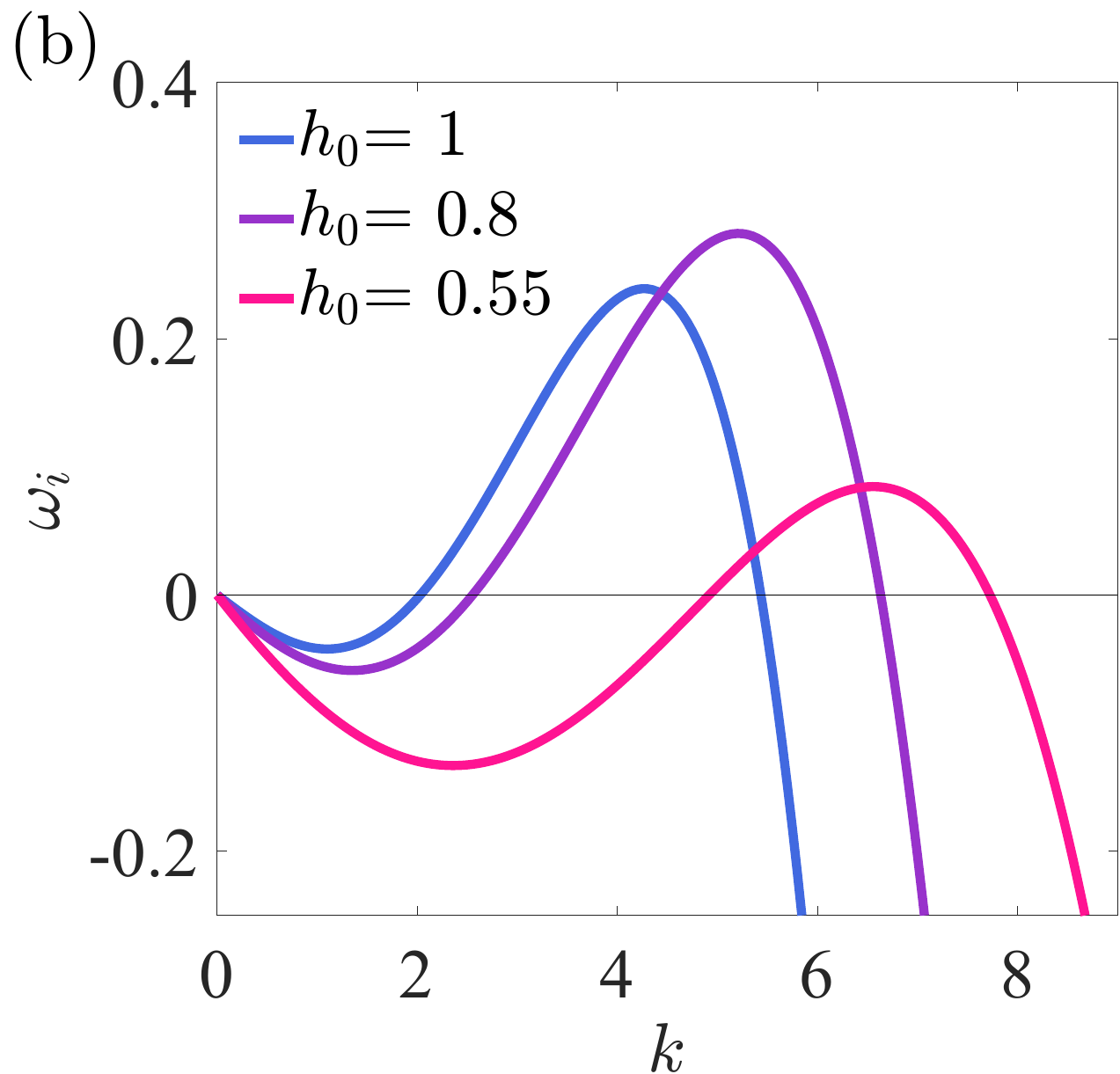}
	\end{subfigure}
	\hskip0.3cm
	\begin{subfigure}[t]{0.3\textwidth}
		\includegraphics[width=\textwidth]
		{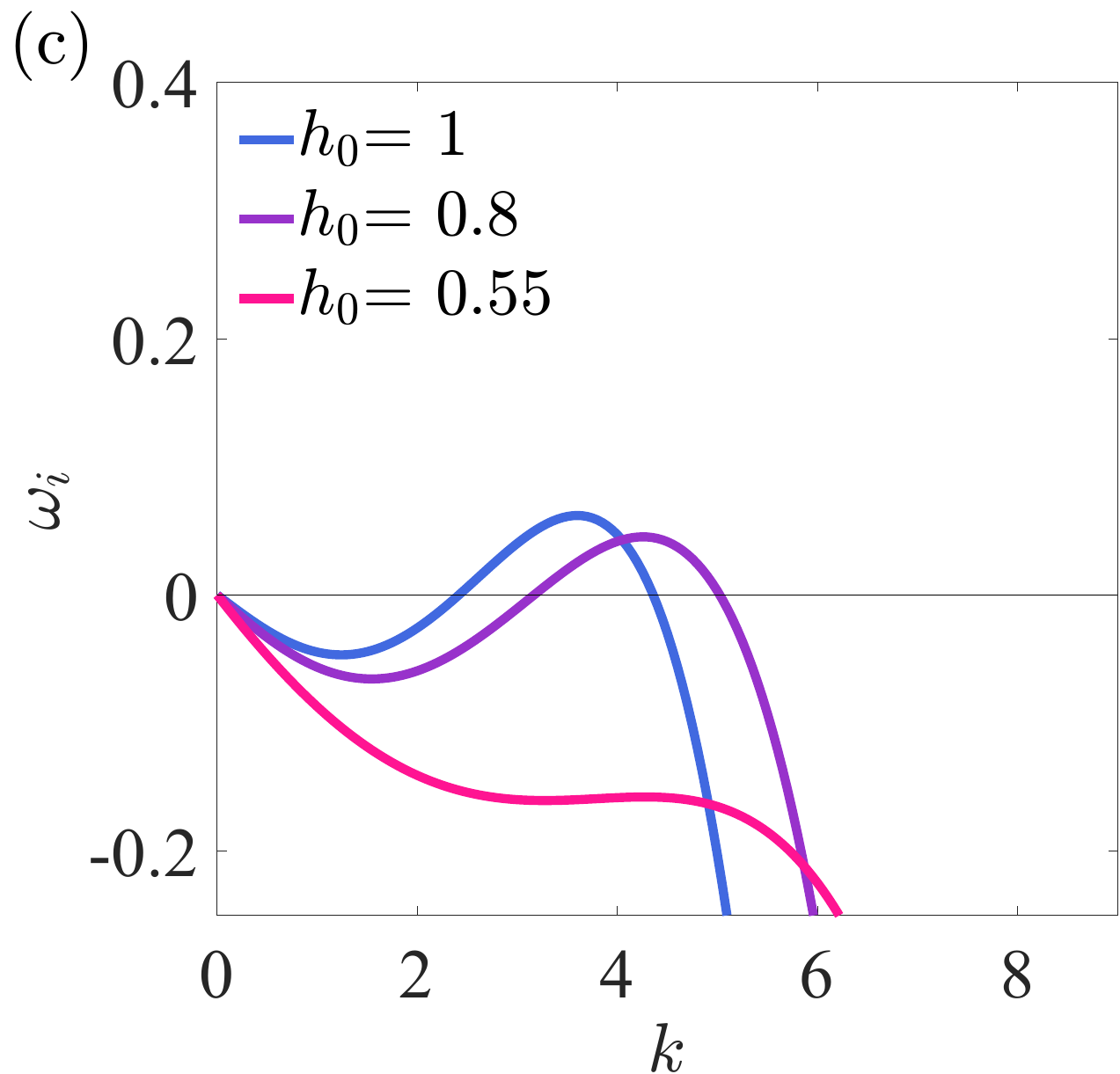}
	\end{subfigure}
	\caption{Temporal growth rates for a thinning film in the diffusion-limited regime $(\Gamma = 1000)$ with varying evaporation intensity. 
	$Re = 2$, $\beta = 15$, $\mathcal{M} = 0.63$, $\bar{E}=0.065$, and
	(a) $V_r = 0.02$,
	(b) $V_r = 0.14$, and
	(c) $V_r = 0.21$.
	}
	\label{temporal_film_height_DF}
\end{figure}

In fact, the effect of the reduction in base flow height on the film's stability in the diffusion-limited regime is rather nuanced and is determined by the relative strength of the evaporation phenomena with respect to the Marangoni instability. This is depicted in Fig. \ref{temporal_film_height_DF} which plots the perturbation's temporal growth rates in the same thinning film for three different strengths of the vapor recoil parameter.
In Fig. \ref{temporal_film_height_DF}(a), vapor recoil is weak with respect to the Marangoni effect $(V_r = 0.02)$. Thus, as the film thins the resulting reduction in the sum of the stabilizing terms is larger than the corresponding decrease in the destabilizing Marangoni term. As a result, the effect of film-thinning is destabilizing.
On the other hand, Fig.\ref{temporal_film_height_DF}(b) depicts the temporal growth rates for the same film where vapor recoil is now significantly stronger $(V_r = 0.14)$. In this case, we witness that the initial reduction in base film height to $\bar{h} = 0.8$ has a destabilizing influence as evidenced by the increase in the maximum temporal growth rate, which is driven by the same destabilizing mechanism observed in Fig. \ref{temporal_film_height_DF}(a). However, as the film height reduces further, the relatively slow $[\sim\mathcal{O}(\bar{h})]$ reduction in the vapor recoil term [see Fig. \ref{film_thinning_growthrate}(d)] signifies that vapor recoil retains its potency as a stabilizing factor relative to the rapidly decaying Marangoni effect. This ultimately leads to an inversion in the impact of film thinning from destabilizing to stabilizing. 
Lastly, when vapor recoil is even stronger, as in Fig. \ref{temporal_film_height_DF}(c), the reduction in base flow height has an immediate stabilizing effect since the decrease in the strength of the Marangoni instability is larger than the decrease in the remaining stabilizing phenomena, which mirrors the case depicted in Fig. \ref{film_thinning_growthrate}(b).
Additionally, we note that due to the relatively weak Marangoni instability, the stabilizing role of mass loss is significant, and consequently, the results in Fig. \ref{temporal_film_height_DF} can be reproduced by increasing the mass loss parameter $\bar{E}$ in place of $V_r$. However, this analogous impact of evaporative mass loss is not depicted here graphically to avoid redundancy.

\subsection{The impacts of evaporation regime and film thinning on wave celerity}\label{phase_speed_section}
\begin{figure}[tp]
	\centering
	\begin{subfigure}[h]{0.355\textwidth}
		\includegraphics[width=\textwidth]
		{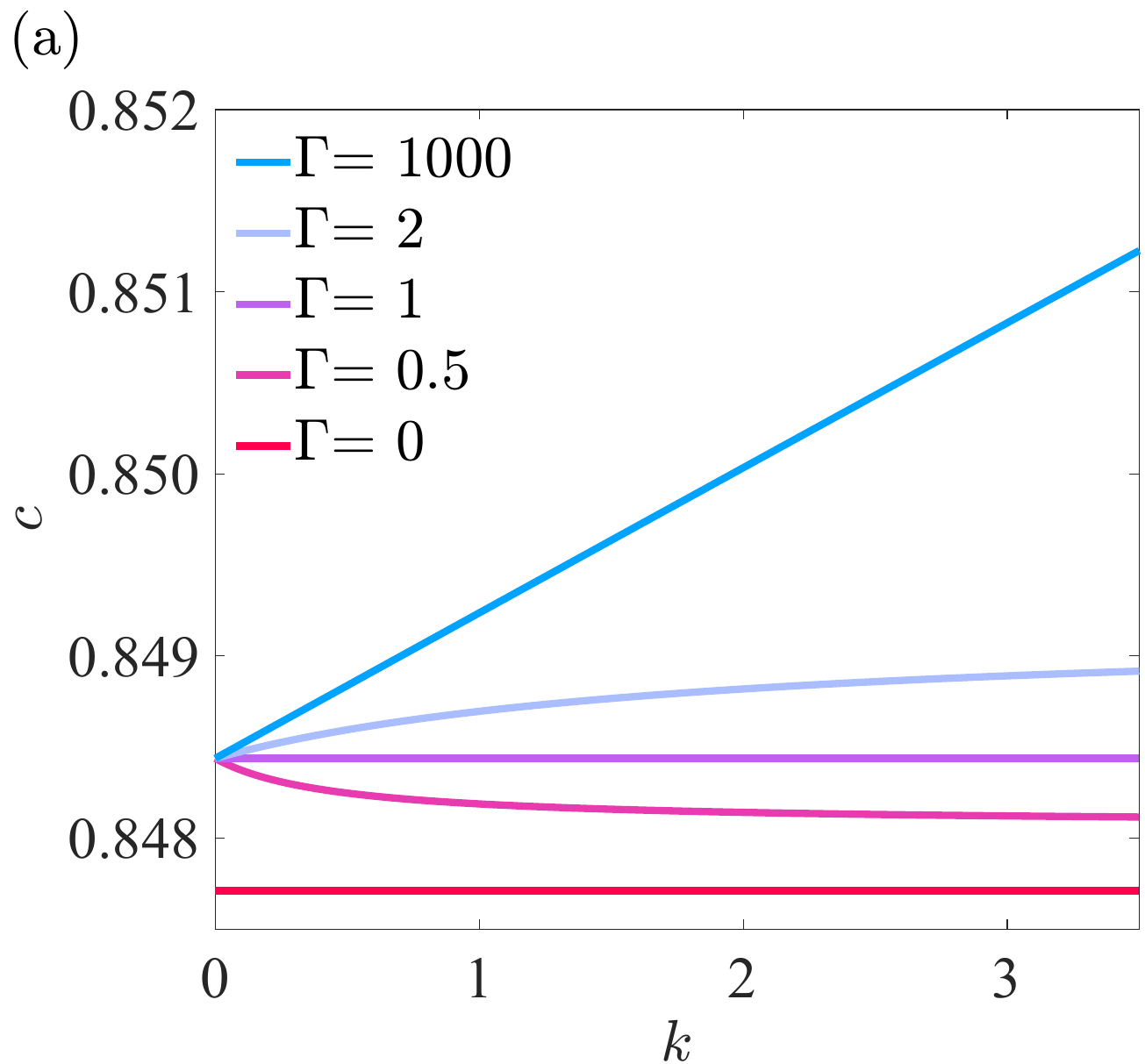}
	\end{subfigure}
	\hskip 1.5cm
	\begin{subfigure}[h]{0.355\textwidth}
		\includegraphics[width=\textwidth]
		{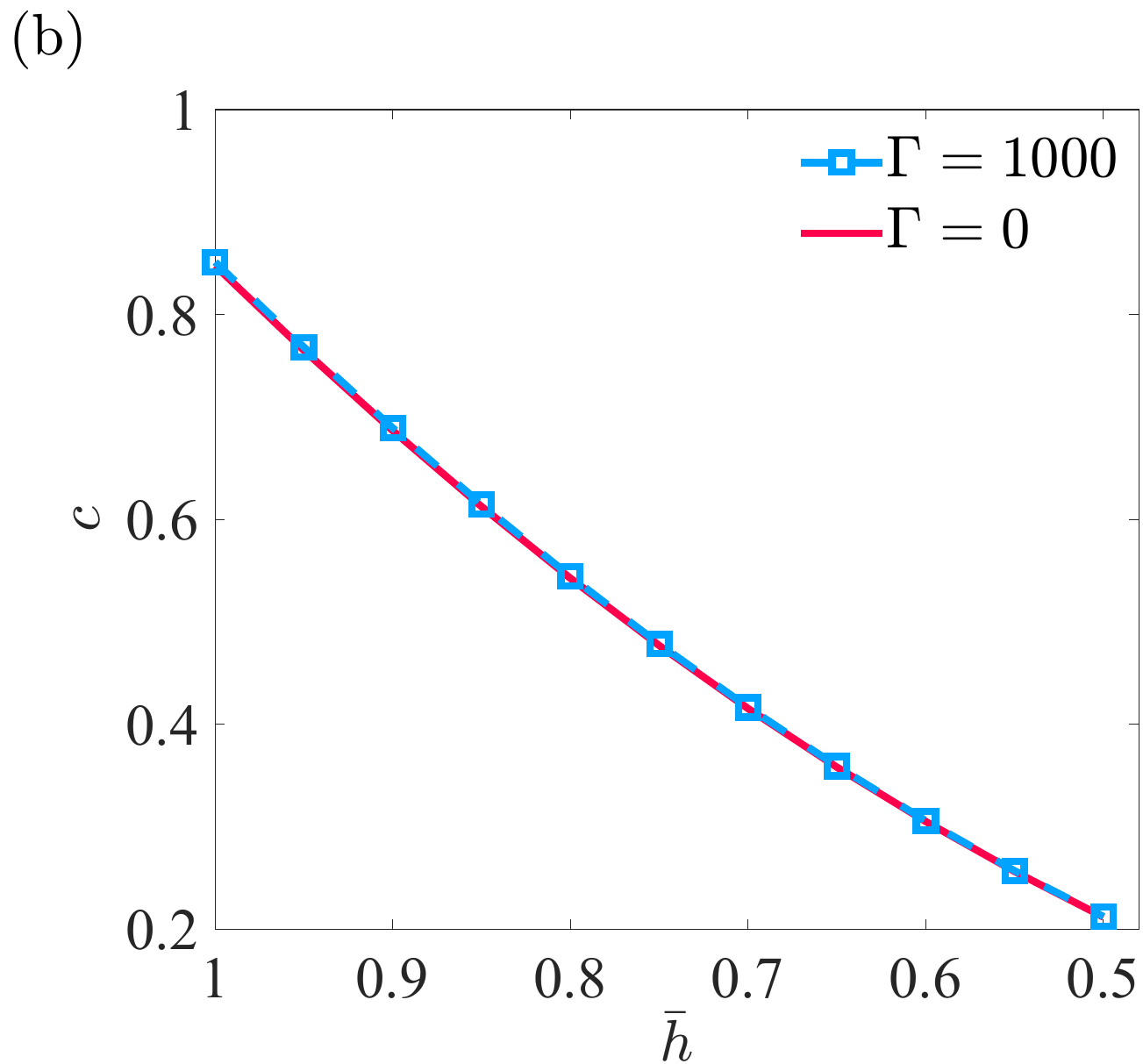}
	\end{subfigure}
	\caption{
		Perturbation phase speed versus (a) wave number and (b) base flow height for
	$Re = 2$, $\beta = 25$, $\mathcal{M} = 1.25$, $V_r = 0.15$, and $\bar{E}=0.1$.
	}
	\label{phase_speed}
\end{figure}

We extend our analysis to investigate the influence of evaporation regime and  film thinning on wave propagation within the film by examining the phase speed $c = \dfrac{\omega_r}{k}$,  which is obtained directly from Eq. \eqref{omega_r_general} as
\begin{align}\label{phase_speed_general}
	c = Re\sin\beta
	\bar{h}^2 + \epsilon\bar{E}\bar{J}Re\sin\beta\left(
	\dfrac{5}{6}\bar{h}^3
	+ \dfrac{\Gamma - 1}{\Gamma + |k|\left(\bar{h} + \dfrac{1}{\chi}\right)}\dfrac{5}{24}\bar{h}^4|k|
	\right).
\end{align}
The last two terms in Eq. \eqref{phase_speed_general} indicate that the phase speed is impacted by evaporative mass loss, mirroring the findings in Refs. \cite{joo1991,mohamed_luca_21}. Intriguingly, the third term is also proportional to $\Gamma$ which implies that the phase speed is affected by the evaporation regime in particular. We demonstrate this dependence in Fig. \ref{phase_speed}(a), which shows the variation in wave dispersion across the evaporation regimes, where we find three distinct dispersion relationships.
For the transfer-rate-limited $(\Gamma \rightarrow 0)$ and balanced flux $(\Gamma = 1)$ cases, the phase speed does not depend on the wave number, and thus, the waves are non-dispersive.
On the other hand, for moderate values of $\Gamma$ representing general evaporation, the waves are nonlinearly dispersive, with this dispersion being positive for $\Gamma < 1$ and negative for $\Gamma > 1$, indicating that the imbalance of vapor diffusion in particular is related to wave dispersion. We note that this positive (negative) dispersion can lead to pulse broadening (compression) when $\Gamma<1$ ($\Gamma>1$). However, the effect is negligible since it is only effective for small $k$. That being said, when $\Gamma=1000$ signifying diffusion-limited evaporation, Eq. \eqref{phase_speed_general} takes the form
\begin{align}\label{phase_speed_DF}
	c = Re\sin\beta
	\bar{h}^2 + \epsilon\bar{E}\bar{J}Re\sin\beta\left(
	\dfrac{5}{6}\bar{h}^3
	+ \dfrac{5}{24}\bar{h}^4|k|
	\right),
\end{align}
which demonstrates that for diffusion-limited evaporation the waves are linearly negatively dispersive for all $k$. This suggests that evaporation effects in the diffusion-limited regime can lead to pulse compression.
Nevertheless, we find that the impact of the evaporation regime on phase speed is entirely overshadowed by that of the film thinning. This is depicted in Fig. \ref{phase_speed}(b), where $c$ is almost identical in both evaporation limits but reduces drastically with $\bar{h}$ due to the stronger viscous resistance, reflecting the findings of Ref. \cite{mohamed_luca_21}.

\section{Spatiotemporal stability analysis}\label{spatiomteporal_stability}
The spatiotemporal dynamics are investigated by determining the flow's long-time response to an infinitesimal impulse source, following the procedure of Huerre and Monkewitz \cite{huerre1990local}. The nature of the instability is determined by examining the temporal growth rate $\sigma(V)$ associated with the zero-group-velocity $(V = 0)$, also known as the absolute growth rate  denoted typically as $\omega_{i,0}$, ($\omega_{i,0}<0$ indicates that the flow is convectively unstable and $\omega_{i,0}>0$ indicates that it is absolutely unstable). As required by our spatiotemporal analysis, we assume both $\omega$ and $k$ to be complex in Eq. \eqref{dispersion_relationship_general}, which is then utilized to obtain the linear impulse response wave packets conveying the nature of the flow.

However, we note that the system's dispersion relationship [Eq. \eqref{dispersion_relationship_general}] in its general form is not analytic at zero due to the presence of $|k|$, which typically prevents the application of the Briggs pinching criterion \cite{huerre1990local}. 
Nevertheless, we can overcome this impediment by restricting our spatiotemporal analysis to wave numbers with positive real part $(|k|\rightarrow +k)$, thereby removing the nonanalyticity from the dispersion relationship. 
This is justified by the fact that the temporal analysis produces a unique maximum temporal growth rate $\omega_{i,max}$ that is also a global maximum for all $\sigma(V)$ \cite{huerre1990local}, and since the expression for the temporal growth rate is symmetric around $k =0$, we expect to see this unique maximum for $|k|\rightarrow \pm k$. Hence, setting $|k|\rightarrow +k$ in our dispersion relationship produces this dominant mode as a saddle point on the right-hand-side of the $(k_r,k_i)$ plane allowing the classification of the instability. Interestingly, the non-analyticicty vanishes from the dispersion relationship [Eq. \eqref{dispersion_relationship_general}] in the transfer-rate-limited $(\Gamma \rightarrow  0)$ and balanced flux $(\Gamma = 1)$ cases, where the vapor gradient does not make an effective contribution to the system.

\begin{figure}[tp]
	\centering
	\includegraphics[width=0.355\textwidth]
	{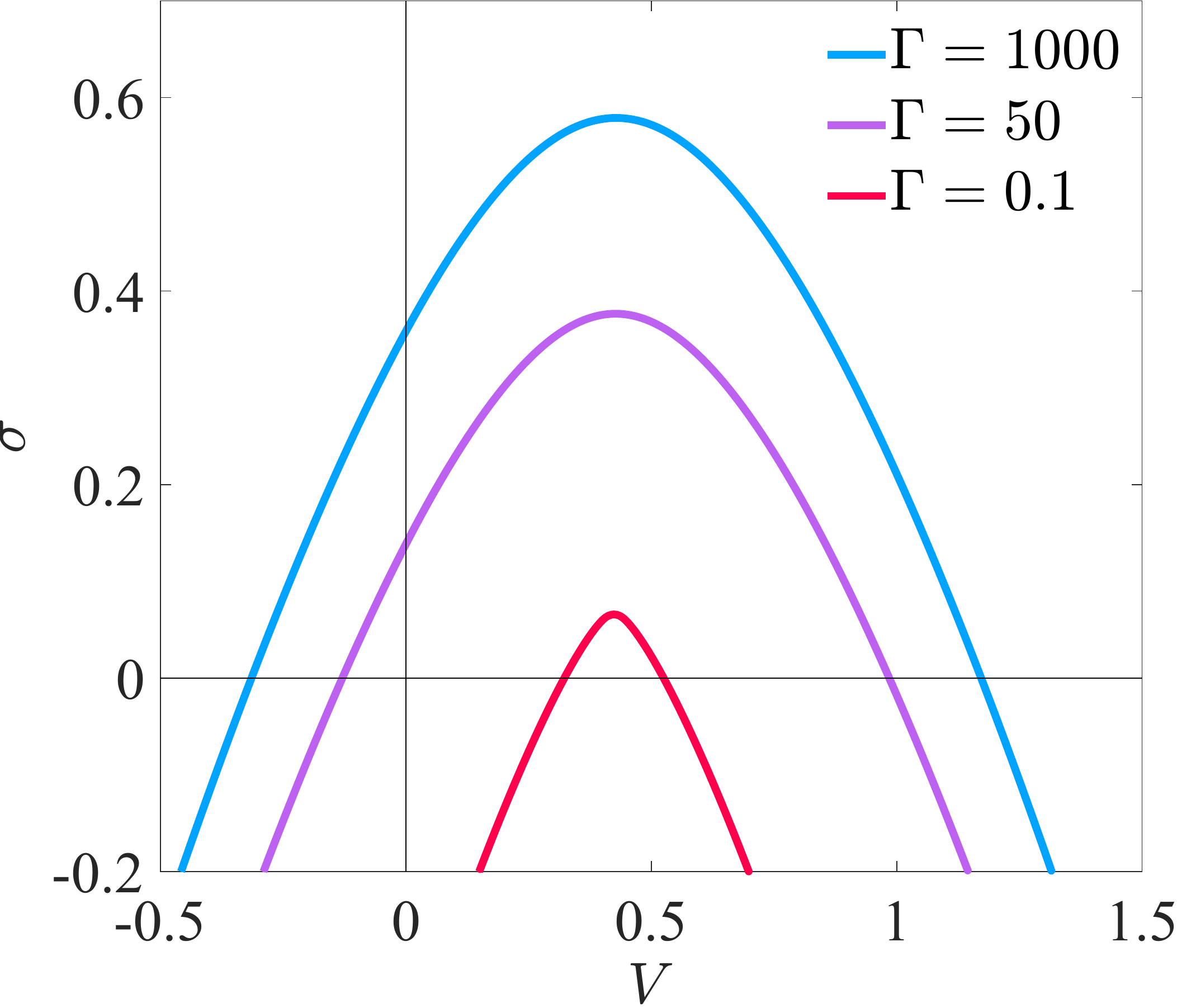}
	\caption{Spatiotemporal impulse responses for $Re = 1$, $\beta = 25$, $\mathcal{M} = 0.7$, $V_r = 0.2$, $\bar{E}=0.1$, 
		and increasing values of $\Gamma$ representing the increasing importance of vapor diffusion.}
	\label{spatiotemporal_evaporation_regime}
\end{figure}

The impact of the evaporation regime on the spatiotamporal stability of the flow can be seen in Fig. \ref{spatiotemporal_evaporation_regime}, where increasing the relative importance of vapor diffusion results in increasing the maximum temporal growth rate as well as expanding the range of unstable modes, as evidenced by the widening of the impulse response wave packets as the value of $\Gamma$ is increased. This is also accompanied by a change in the spatiotemporal nature of the flow from convectively unstable in the $\Gamma = 0.1$ case where vapor diffusion is relatively insignificant, to absolutely unstable for $\Gamma=50$ which represents a case of evaporation dominated by vapor diffusion. Increasing the value of $\Gamma$ even further leads to a corresponding increase in the absolute growth rate, as seen in the case of $\Gamma = 1000$ representing the diffusion-limited regime. 
The spatiotemporal analyis reveals that the evaporation regime has a significant effect on the propagation of the perturbations in the film. This effect is however quite complex, and cannot be detected by simple observation of the phase speed in the temporal framework [See Fig. \ref{phase_speed}(b)]. \color{black}
\begin{figure}[tp]
	\centering
	\includegraphics[width=0.355\textwidth]
	{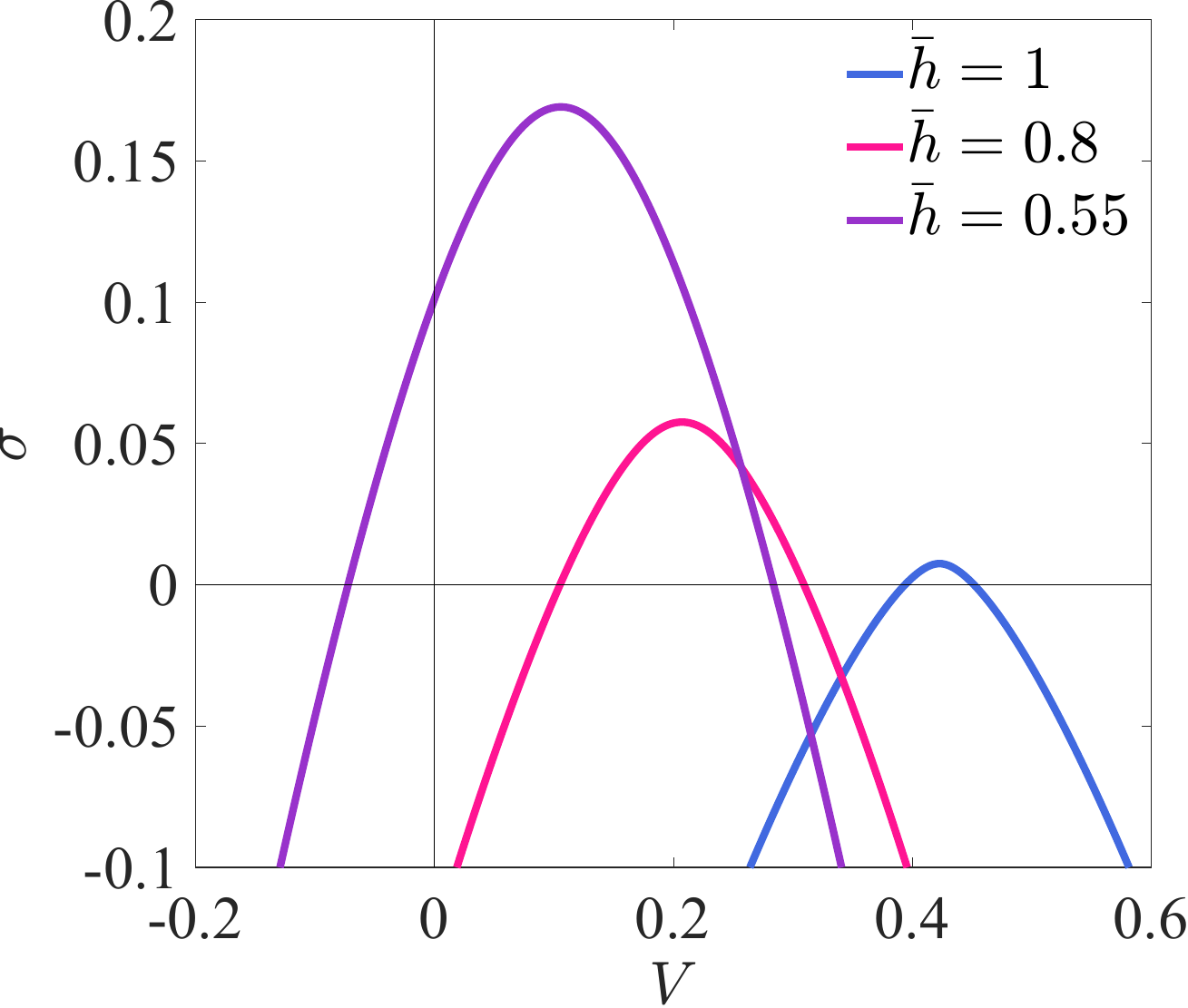}
	\caption{Spatiotemporal impulse responses for a thinning film in the transfer-rate-limited regime $(\Gamma = 0)$ for
		$Re = 2$, $\beta = 15$, $\mathcal{M} = 2.75$, and $V_r = 0.35$, $\bar{E}=0$.
	}
	\label{spatiotemporal_film_height_TL}
\end{figure}
Moreover, the evaporation regime also determines the  influence of the film thinning on the spatiotemporal stability of the flow, as shown in Fig. \ref{spatiotemporal_film_height_TL}. In the transfer-rate-limited regime $(\Gamma \rightarrow 0)$, the reduction in base flow height results in increasing both the maximum temporal growth rate and the absolute growth rate which can lead to transitions in the spatiotemporal nature of the instability. Initially the flow is convectively unstable at $\bar{h}=1$ as both edges of the impulse response are on the same side of the vertical axis. As the film becomes thinner, the maximum temporal growth rate is increased and the convective nature of the instability is weakened, as evidenced by the impulse response shifting closer to the origin resulting in a higher value of $\omega_{i,0}$. Further thinning of the film height leads to additional destabilization as well as causing the left edge of the impulse response to cross the $y$-axis (i.e., $V=0$) above the origin, producing a positive value of $\omega_{i,0}$. This signifies that the flow has undergone a transition from convective to absolute instability. Note that this behavior matches the findings reported in Ref. \cite{mohamed_luca_21} on the effect of film thinning on spatiotemporal stability in the transfer-rate-limited regime. 
\begin{figure}[tp]
	\centering
	\begin{subfigure}[t]{0.3\textwidth}
		\includegraphics[width=\textwidth]
		{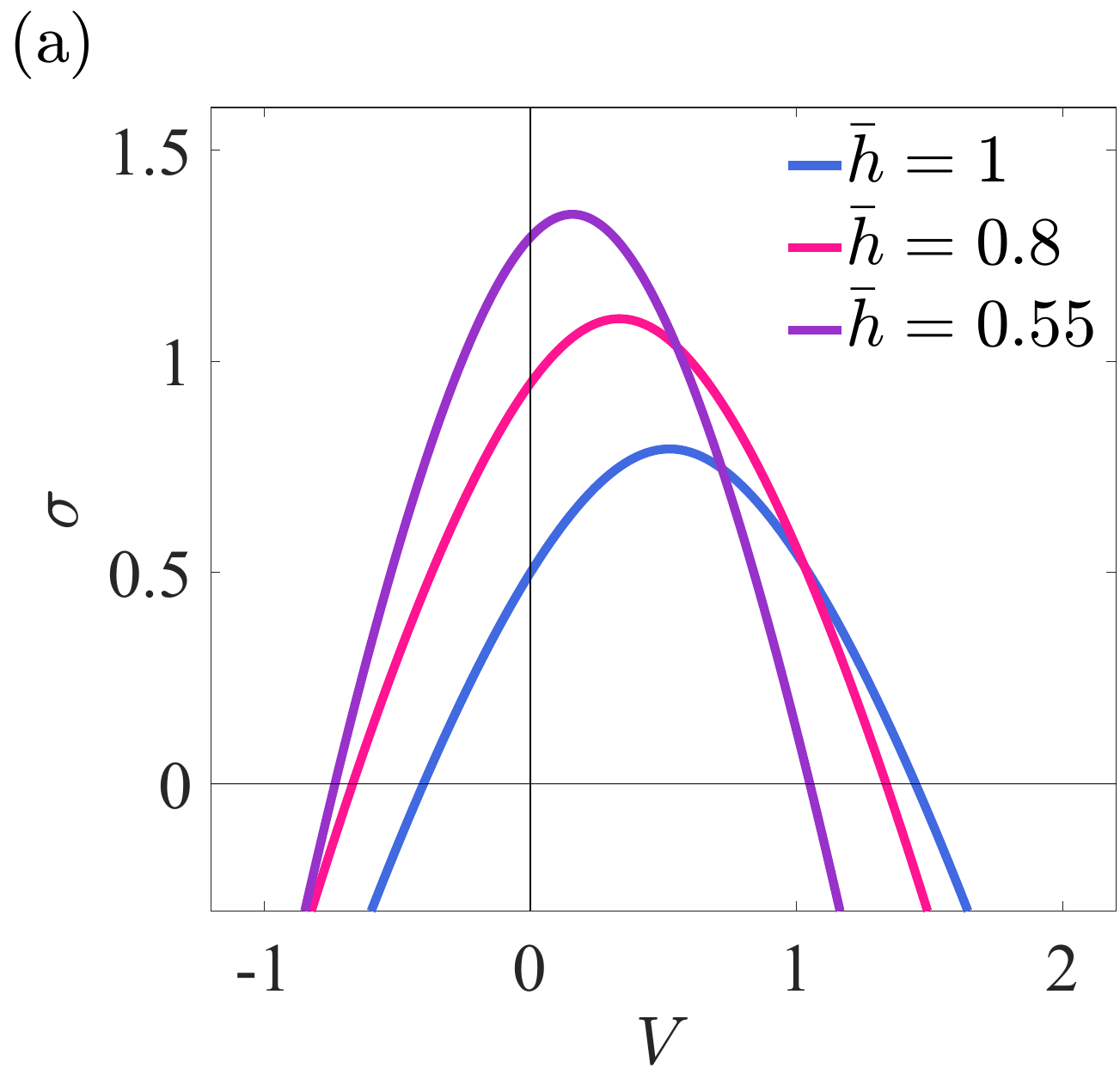}
	\end{subfigure}
	\hfill
	\begin{subfigure}[t]{0.3\textwidth}
		\includegraphics[width=\textwidth]
		{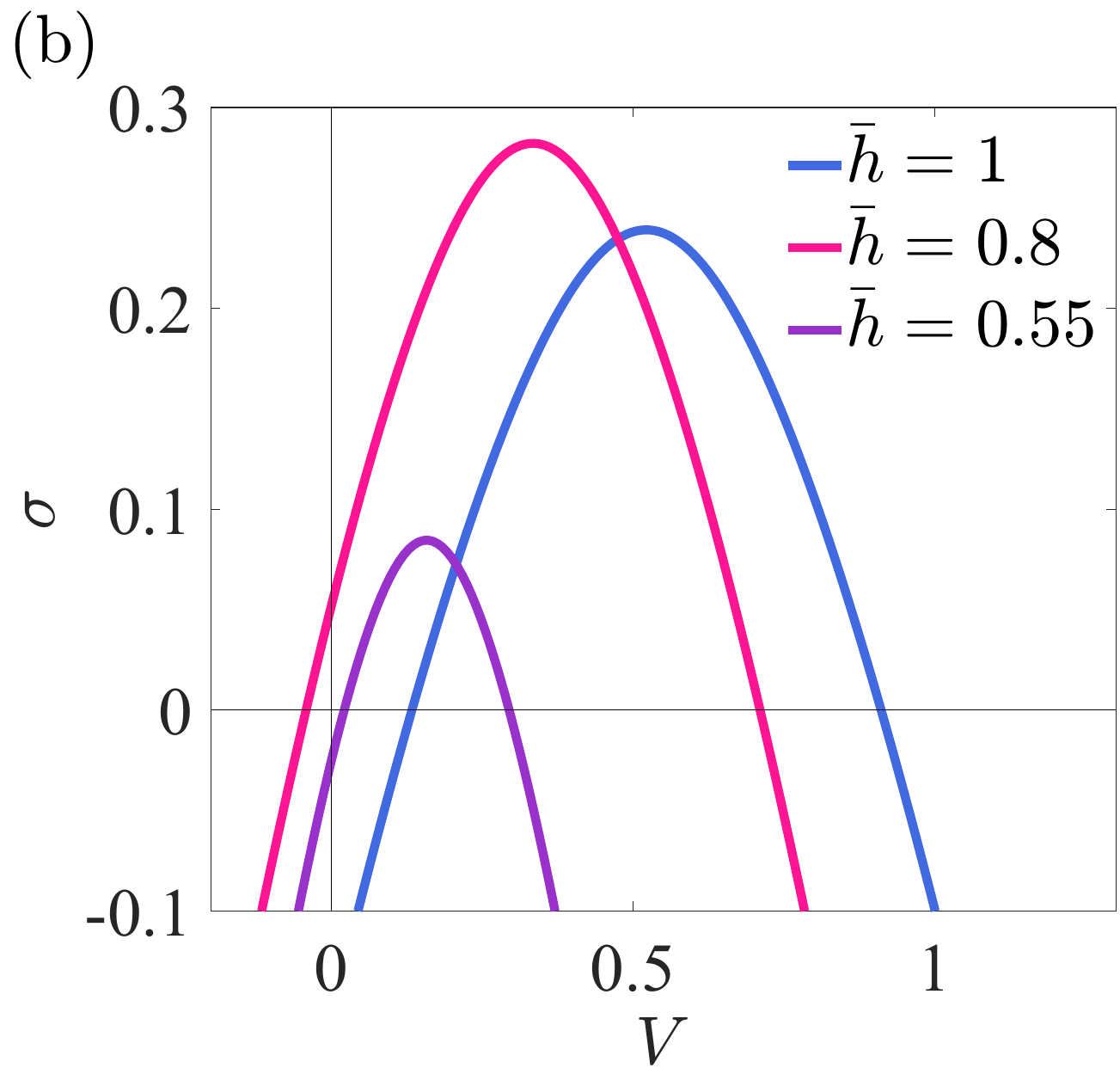}
	\end{subfigure}1
	\hfill
	\begin{subfigure}[t]{0.3\textwidth}
		\includegraphics[width=\textwidth]
		{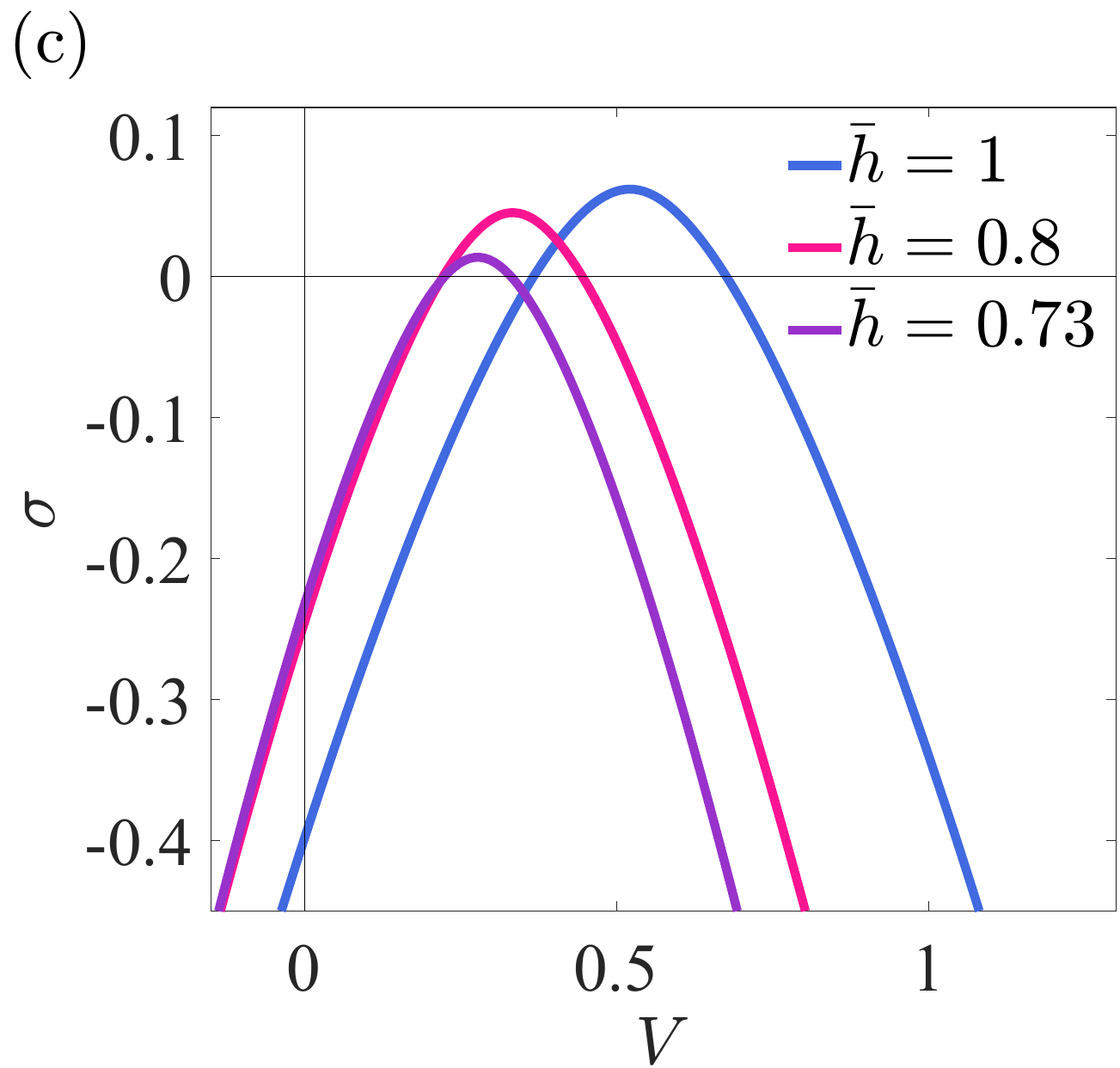}
	\end{subfigure}
	\caption{Spatiotemporal impulse responses for a thinning film in the diffusion-limited regime $(\Gamma = 1000)$ with varying evaporation intensity. 
		$Re = 2$, $\beta = 15$, $\mathcal{M} = 0.63$, $\bar{E}=0.065$, and
		(a) $V_r = 0.02$,
		(b) $V_r = 0.14$, and
		(c) $V_r = 0.21$.
	}
	\label{spatiotemporal_film_height_all_VR}
\end{figure}

On the other hand, the impact of film thinning on spatiotemporal stability in the diffusion-limited regime is quite complex and emerges from the rich interactions between the Marangoni and evaporation phenomena, where the intensity of evaporation plays a fundamental role in determining the spatiotemporal nature of the flow. We illustrate this by conducting a spatiotemporal analysis with the same parameter sets used in the temporal analysis (see Fig. \ref{temporal_film_height_DF}), where the impact of stronger evaporation is again represented by increasing the value of $V_r$.
We observe in Fig. \ref{spatiotemporal_film_height_all_VR}(a), where the vapor recoil is relatively weak $(V_r = 0.02)$, that the reduction in film thickness has an effect akin to that in the transfer-limited case depicted in Fig. \ref{spatiotemporal_film_height_TL}. 
Furthermore, Fig. \ref{spatiotemporal_film_height_all_VR}(b) corresponds to Fig. \ref{temporal_film_height_DF}(b), where vapor recoil is significantly stronger. The influence of film thinning in this intermediate case is non-monotonous, as we notice that the instability is at first of the convective type for $\bar{h} =1$, and as expected, the reduction in base flow height is initially destabilizing. Notably, this destabilization is accompanied by a transition from convective to absolute instability for $\bar{h} = 0.8$. However, as the base flow height decreases further, film thinning converts into a stabilizing influence and the instability reverts back to the convective type. 
Additional increases in vapor recoil intensity restore the monotonous impact of film thinning where this effect is immediately stabilizing, as indicated by the progressive decrease in $\omega_{i,max}$ with $\bar{h}$ in  Fig. \ref{spatiotemporal_film_height_all_VR}(c). 
Nevertheless, the reduction in base flow height in this case still results in an increase in the absolute growth rate, subduing the flow's convective character. 

This propensity of the reduction in film height to shift the impulse response towards the vertical axis is observed in all three cases of Fig. \ref{spatiotemporal_film_height_all_VR}, and is tied to the inverse relationship between phase speed and film height [see Fig. \ref{phase_speed}(b)], since the center of the impulse response wave packet corresponds to $c(k_{max})$. We note again that this dynamic interplay between film thinning and vapor recoil strength can be replicated by increasing the intensity of the mass loss effect i.e., by raising the value of $\bar{E}$, however, it is not depicted here to avoid redundancy.

\section{Conclusions}\label{conclusion}
We have investigated the linear stability of an evaporating liquid film flowing down a inclined plane across a wide range of evaporation regimes by theoretical analysis and numerical simulation. We modified the general evaporation model of Sultan \textit{et al.} \cite{sultan2005evaporation} to allow for thermodynamic disequilibrium and the resultant mass loss, and combined it with the film model of Joo \textit{et al.} \cite{joo1991} to obtain a coupled liquid-vapor system. Subsequently, we performed linear stability analyses in both the temporal and spatiotemporal frameworks.
The main focus of our investigation was the impact of the evaporation regime on the Marangoni instability and its complex coupling to the evaporation effects, namely, vapor recoil, mass loss, and film thinning. These rich interactions were yet to be reviewed comprehensively in the presence of vapor diffusion effects; which was the inspiration behind this work. \color{black}

Under our formulation, the Marangoni effect has two components: the first of which is linked to the heat flux rising through the bulk of the liquid film, is dependent only on the liquid interface's curvature, and is always destabilizing.
The second component on the other hand, is caused by the imbalance in vapor diffusion above the interface, and is influenced by both the liquid interface's curvature and the vapor gradient above it. The specific contribution of this component depends on the evaporation regime: it is stabilizing and causes an attenuation of the Marangoni instability when evaporation is dictated by kinetics, while it is highly destabilizing and significantly strengthens the Marangoni instability when vapor diffusion is the predominant effect.
This dependence on the evaporation regime is also shared by the mass loss and vapor recoil effects, which are destabilizing when evaporation is dominated by kinetics, and stabilizing when it is governed by diffusion.

The presence of the diffusion-dependent component of the Marangoni instability significantly destabilizes the film due to the heightened latent cooling imbalance introduced by the vapor density gradient, with this effect being maximized under diffusion-limited evaporation. 
Nevertheless, this destabilizing mechanism competes against vapor recoil and evaporative mass loss, which play a stabilizing role in diffusion-dominated evaporation. These evaporative effects when sufficiently strong can completely subdue the Marangoni instability and restore the stable film.
Interestingly, when the influences of kinetic and diffusion phenomena on the evaporation process are perfectly balanced, the vapor mass flux is equalized across the film which negates the vapor recoil and mass loss instabilities. We refer to this specific scenario in the evaporation regime as the \textit{balanced flux} case.

Moreover, we investigated the impact of the evaporation regime on the mass loss instability and elucidated its relationship to the wave number. We attribute the independence of the mass loss's contribution to the temporal growth rate from the wave number in the transfer-rate-limited regime to the absence of a vapor density gradient above the liquid film, which allows the vapor molecules to freely depart the liquid interface, and leaves the local film height as the sole variable dictating the instability. 
On the other hand, accounting for the liquid's vapor introduces a variable density gradient reliant on the interface's shape, which regulates the molecules' propensity to depart from it and connects the instability to the wave number. 

The influence of film thinning on the film's stability is also investigated at the limits of the evaporation regime. In the transfer-rate-limited regime, film thinning destabilizes the film since the decrease in the strength of the stabilizing forces with the film's height far outpaces the weakening of the destabilizing phenomena.
Conversely, the influence of film thinning in the diffusion-limited regime is quite nuanced and depends on the relative strength of the evaporation phenomena to the Marangoni instability. In fact, the effect varies between being (i) destabilizing when evaporation effects are relatively weak, being (ii) nonmonotonic when for moderately strong evaporation, and (iii) stabilizing when evaporation effects are on par in strength with the Marangoni instability.


Finally, our spatiotemporal analysis revealed that the increase in the temporal growth rate with the rising importance of vapor diffusion is accompanied by an increase in the absolute growth rate, representing a shift towards absolute instability and leading to convective-absolute transitions. 
We also investigated the impact of film thinning on the spatiomteporal stability, where we witnessed a progression towards absolute instability as the film became thinner in the transfer-rate-limited regime, and an intricate relationship between evaporation effects and the absolute growth rate in the diffusion-limited regime. Under relatively weak evaporation, the decrease in film height increases the absolute growth rate and brings forth convective absolute transitions, whereas moderate evaporation brings about nonmonotonic changes in the absolute growth rate leading to convective-absolute-convective transitions as the film thins. Nevertheless, relatively strong evaporation restores the monotonic increase in the absolute growth rate as the film thins.

Accounting for vapor diffusion in the stability analysis of an evaporating liquid film gives rise to several complex and interesting dynamics resulting from the competition between the various physical phenomena at the liquid interface.
Nevertheless, this problem still offers plenty of opportunities for further study beyond our current investigation.
The limitations imposed by the long wave expansion can be relaxed by formulating the Orr-Sommerfeld eigenvalue problem while including the kinetic-diffusion balance at the liquid interface, which would allow reliably investigating perturbations at higher wave numbers. The results obtained from such models could then be compared to those obtained from numerical simulations of the two-phase problem \cite{LucaBradntEvap2020}. Moreover, more advanced evaporation models that capture additional effects could be used such as that of Zhao and Nadal \cite{Zhao_Nadal_2023}, which accounts for accounts for the Stefan ﬂow, the temperature jump across the Knudsen layer, and the change in the interface's chemical potential due to the presence of the inert gas. Furthermore, the effect of vapor diffusion on other instabilities such as the transverse Marangoni \textit{P}-mode \cite{goussis1991surface} could be studied. Additionally, three-dimensional effects such as boundary confinement can be included as they have been shown to significantly alter the stability of liquid films \cite{vellingiri2015absolute,Mohamed_Sesterhenn_Biancofiore_2023}.

\section*{Acknowledgments}
L.B. would like to acknowledge the Turkish National Research Agency (TÜBİTAK) for supporting this study through Project No. 221M500, which contributed significantly towards the completion of this investigation.

\appendix
\section{NUMERICAL SOLUTION PROCEDURE AND VALIDATION}\label{appendix_numerical}
The liquid-vapor problem is simulated by numerically solving the system formed by the coupling of Eqs. \eqref{vapor_system_ND} and Eq. \eqref{Benney_equation} for a perturbed initial interface described by $h(x,0) = \bar{h}_0 + H\sin(kx)$, where $\bar{h}_0$ is the initial flat film height, and $H$ is the infinitesimal initial perturbation amplitude. The vapor density problem  described by Eqs. \eqref{vapor_system_ND} is solved by means of a finite difference approach utilizing the iterative successive over-relaxation method \cite{young1954iterative}. Due to the interface's curvature, the Cartesian $(x,y)$ coordinate system must be transformed into a boundary-fitted coordinate system $(\xi,\eta)$ using
\begin{align}\label{coordinate_mapping}
	\xi = x, \qquad		\eta = \dfrac{y - y^*_{max}}{h(x,t) - y^*_{max}},
\end{align}
where the mapping in Eq. \eqref{coordinate_mapping} guarantees a uniform computational plane with constant $\Delta\xi$ and $\Delta\eta$ as required by the finite difference approach. This remapping drastically complicates the spatial derivatives in the system, but results in one-to-one correspondence between the solution points in the computational and physical spaces.
For further details on grid transformations, please see Ref. \cite{Anderson1995}.
%

\begin{figure}[tp]
	\centering
	\includegraphics[width=0.31\textwidth]	
    {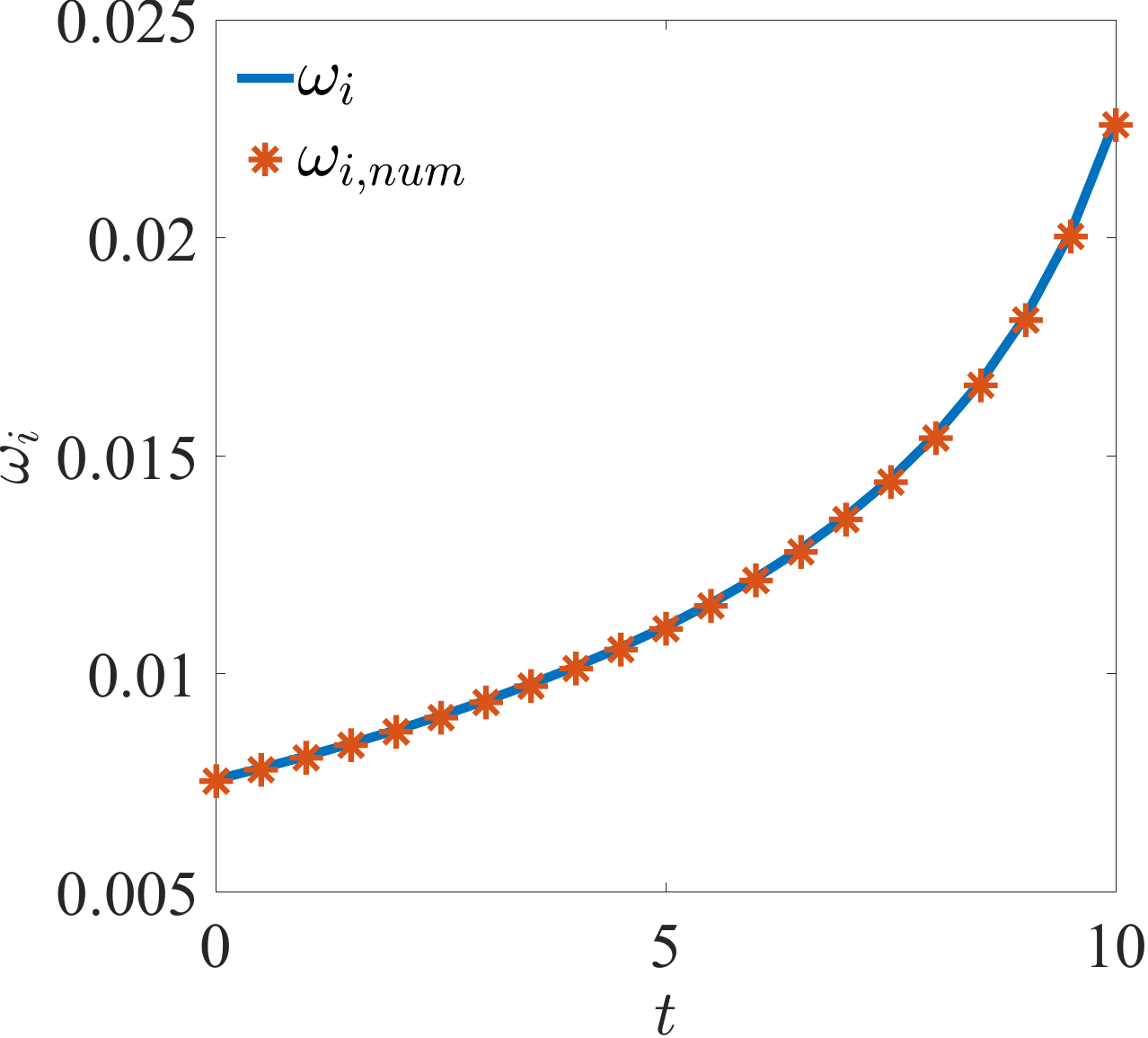}
	\caption{Comparison of the analytical and numerical temporal growth rates for 		
		$Re = 2$, $\beta = 25$, $\mathcal{M} = 0.75$, and $V_r = 0.15$, $\bar{E}=0.05$, $\Gamma = 0.5$, and a grid size of $(N_x,N_y) = (32,2048)$.
		The initial interface in the numerical simulation is described by $ h(x,0) = 1 + 0.001\sin(0.25x)$. The difference between $\omega_i$ and $\omega_{i,num}$ remains less than $0.6\%$.}
	\label{numerical_validation}
\end{figure}

The evaporative mass flux from the liquid interface can be computed efficiently by rearranging the interfacial boundary condition [Eq. \eqref{BC_vapor_ND}] as:
\begin{align}
	J = - \gamma\dfrac{\rho_{v|int}}{h + 1/\chi}\Gamma.
\end{align}
$J$ is then substituted into the liquid film's evolution equation [Eq. \eqref{Benney_equation}], which is integrated as an initial value problem using MATLAB \cite{MATLAB2023b}.
The numerical temporal growth rate is extracted by performing the discrete Fourier transform of $h(x,t)$ at each time step as
\begin{subequations}
\begin{equation}
	\hat{f}(k,t) = \sum_{j = 1}^{n-1}h(j,t)e^{-2\pi i(j - 1)(k - 1)/(n-1)},
\end{equation}
where $n$ is the total number of discrete points in $h(x,t)$. The perturbation amplitude at each time step is then found from
\begin{equation}
	\hat{H}(k,t) = 2|\hat{f}(k,t)|, 
\end{equation}	
which allows computing the numerical temporal growth rate as
\begin{equation}
	\omega_{i,num} = \dfrac{\dot{\hat{H}}}{\hat{H}}.
\end{equation}
\end{subequations}
Our numerical solver and analytical derivation of the temporal growth rate [Eq. \eqref{omega_i_general}] are cross-validated by comparing $\omega_i$ and $\omega_{i,num}$ as shown in Fig. \ref{numerical_validation}, where the difference between the numerical  analytical growth rates does not exceed $0.6\%$.

Note that the number of grid points along the $y$ coordinate is varied according to strength of the vapor density gradient across the vapor domain, as represented by the value of $\Gamma$. For very small values of $\Gamma$, the vapor density is approximately constant and therefore, a relatively small number of grid points is sufficient to obtain accurate simulation results. On the other hand, relatively large values of $\Gamma$ are associated with sharp density gradients across the vapor domain requiring a comparatively large number of grid points to accurately resolve. In practice, we varied the number of grid points along the $y$-direction from $N_y=8$ to $N_y=2048$ depending on the value of $\Gamma$. Moreover, in all the simulations presented, we utilized a maximum vapor domain height $y_{max} = 35$, which is sufficiently large to accurately model our system while retaining numerical efficiency.

\bibliography{refs}

\end{document}